\documentclass[aps,twocolumn,english,superscriptaddress,citeautoscript,preprintnumbers,amsmath,amssymb,floatfix,footinbib]{revtex4-1}
\usepackage{hyperref}
\hypersetup{colorlinks=true,linkcolor=red,citecolor=blue}
\usepackage{amsmath}
\usepackage{graphicx}
\usepackage{dcolumn}
\usepackage{float}

\begin{document}

\title{Evolving Devil's staircase magnetization from tunable charge density waves in nonsymmorphic Dirac semimetals}

\author{Ratnadwip Singha}

\affiliation{Department of Chemistry, Princeton University, Princeton, New Jersey 08544, USA}

\author{Tyger H. Salters}

\affiliation{Department of Chemistry, Princeton University, Princeton, New Jersey 08544, USA}

\author{Samuel M. L. Teicher}

\affiliation{Materials Department and Materials Research Laboratory, University of California, Santa Barbara, California 93106, USA}

\author{Shiming Lei}

\affiliation{Department of Chemistry, Princeton University, Princeton, New Jersey 08544, USA}

\author{Jason F. Khoury}

\affiliation{Department of Chemistry, Princeton University, Princeton, New Jersey 08544, USA}

\author{N. Phuan Ong}

\affiliation{Department of Physics, Princeton University, Princeton, New Jersey 08544, USA}

\author{Leslie M. Schoop}

\email{lschoop@princeton.edu}

\affiliation{Department of Chemistry, Princeton University, Princeton, New Jersey 08544, USA}


\begin{abstract}

While several magnetic topological semimetals have been discovered in recent years, their band structures are far from ideal, often obscured by trivial bands at the Fermi energy. Square-net materials with clean, linearly dispersing bands show potential to circumvent this issue. CeSbTe, a square-net material, features multiple magnetic field-controllable topological phases. Here, it is shown that in this material, even higher degrees of tunability can be achieved by changing the electron count at the square-net motif. Increased electron filling results in structural distortion and formation of charge density waves (CDWs). The modulation wave-vector evolves continuously leading to a region of multiple discrete CDWs and a corresponding complex ``Devil's staircase'' magnetic ground state. A series of fractionally quantized magnetization plateaus are observed, which implies direct coupling between CDW and a collective spin-excitation. It is further shown that the CDW creates a robust idealized non-symmorphic Dirac semimetal, thus providing access to topological systems with rich magnetism.

\end{abstract}

\maketitle

\section{Introduction}

The quest for finding novel quantum phases drives the discovery of new materials and phenomena in condensed matter physics. The introduction of topological band structure in material classification, is perhaps the most notable example in the last decade. Starting from the identification of the topological insulating state in a handful of materials \cite{Hsieh, Chen, Xia, Hsieh2}, this field has been enriched by subsequent realization of a plethora of unconventional quantum states such as Dirac \cite{Liu, Liu2}, Weyl \cite{Lv, Xu, Xu2}, nodal-line semimetals \cite{Bian, Schoop2}, and topological superconductors \cite{Beenakker}. While the idea of exploring the dynamics of relativistic particles in low-energy electronic systems is already quite enticing, the prospect of probing new quasiparticle excitations beyond the realm of particle physics \cite{Bradlyn} is even more attractive. Motivated by these predictions, researchers have utilized data mining to identify and categorize a huge number of compounds hosting different types of non-trivial topological band structures \cite{Bradlyn2, Vergniory}. To realize new topological phases, another effective route is to introduce additional tuning parameters in these materials such as strong electron correlations \cite{Lu, Kondo} and magnetism. Incorporation of magnetism in topological systems is particularly interesting as it can break the inherent time-reversal symmetry (TRS) of the crystal structure. TRS broken topological semimetals are extremely rare and have been the major focus of a number of studies \cite{Kuroda, Liu3, Belopolski, Schroter}.\\

Contrary to the approach of scanning through material databases, it has been shown that simple structural motifs can provide strong indications of topological non-trivial states in a compound \cite{Schoop}. A great example of this are square-net materials. In 2015, Young and Kane proposed nonsymmorphic symmetry protected Dirac semimetal states in a two-dimensional (2D) square-net lattice \cite{Young}. In recent times, a vast array of topological semimetals have been discovered, all hosting the 2D square-net motif in their crystal structure \cite{Klemenz}. Probably the most extensively studied member of this family is ZrSiS, which shows the largest reported energy range ($\sim$2 eV) of linearly dispersive bands \cite{Schoop2}. It hosts multiple Dirac nodes near the Fermi energy, which form a diamond shaped nodal-line in momentum space \cite{Schoop2, Singha, Ali}. In addition, both far below and above the Fermi energy, there are nonsymmorphic symmetry protected Dirac nodes, which, in contrast to other Dirac cones, remain gapless even under spin-orbit coupling \cite{Schoop2}. Thus, ZrSiS presents an unique platform to explore different types of topological physics. CeSbTe is an isostructural compound, which in addition to having a rich band structure similar to ZrSiS, introduces magnetism as an additional controllable parameter \cite{Schoop3}. In fact, it has been reported that this material can be driven through distinct topological phases while changing the magnetic ordering by modulating the strength of an applied magnetic field, resulting in a tunable topological Dirac/Weyl semimetal \cite{Schoop3}. Therefore, CeSbTe is an ideal template to investigate the interplay of magnetism and different topological phases, accessible via Fermi level engineering by electron filling \cite{Weiland}.\\

It is well established that square-net lattices are inherently unstable and only preserved by delocalized electrons as in graphene \cite{Tremel}. Changing the electron count in the square-net, therefore, induces distortion in the crystal structure \cite{Tremel, Papoian}. Such a distorted lattice supports the formation of charge density waves (CDWs), which open a band gap at the Fermi level. In a few rare earth antimony tellurides ($Ln$SbTe; $Ln$=lanthanides), the presence of CDWs has been confirmed when there is a deviation from nominal stoichiometry. For example, in LaSb$_{x}$Te$_{2-x}$, the CDW has been observed to evolve continuously with antimony substitution \cite{DiMasi}. On the other hand, in GdSb$_{x}$Te$_{2-x}$, the CDW is shown to have an important role in designing new Dirac semimetals \cite{Lei}. Recently, the signature of a five-fold CDW modulation has also been reported in an off-stoichiometric single crystal of CeSbTe \cite{Li}. However, the detailed evolution of this phase and its impact on the topological band structure remains unexplored. Such an investigation is important to reveal the true potential of doped $Ln$SbTe, in respect to their magnetic, structural, and topological properties as well as their interplay.\\

In this report, we combine structural and magnetic measurements with theoretical calculations to probe the interplay of magnetic ordering, CDW, and topological band structure in single crystalline CeSb$_{x}$Te$_{2-x-\delta}$ ($\delta$ represents the vacancy concentration in the crystal). We show that there are two distinct regimes in this material with formation of either one or multiple CDW orderings. Moreover, it is possible to continuously tune the modulation wave-vector by controlling the electron filling at the square-net position. The corresponding unique distortions of the square-net motif lead to modifications in the electronic band structure. While CeSbTe has a simple antiferromagnetic type-A ground state, a complex magnetic structure starts to emerge with chemical substitution. In particular, for a certain electron filling range, fractionally quantized magnetization in the form of a ``Devil's staircase'' is observed. We explain the origin of this phenomenon as the consequence of direct coupling between the CDW and a collective spin excitation. From first-principle calculations, we confirm that electron filling and the CDW formation result in an idealized magnetic Dirac semimetal with a nonsymmorphic, symmetry-protected Dirac node at the Fermi energy. This leads to the opportunity to study the interplay of complex magnetism with clean Dirac nodes in topological systems.\\

\begin{figure}
\includegraphics[width=0.45\textwidth]{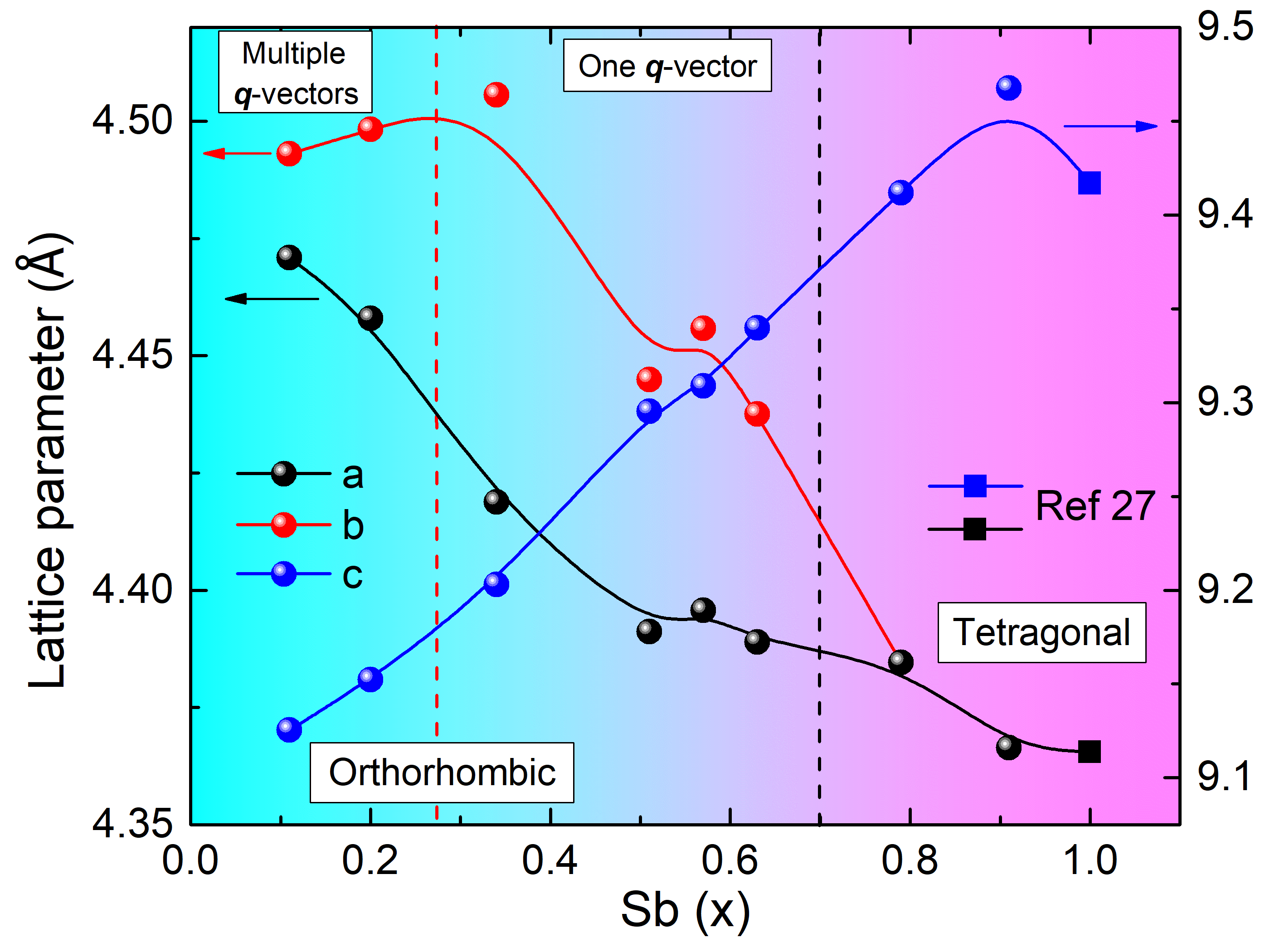}
\caption{Powder x-ray diffraction measurements on CeSb$_{x}$Te$_{2-x-\delta}$ crystals. Lattice constants extracted from powder x-ray diffraction measured on ground single crystals as a function of Sb-content. The dashed black vertical line corresponds to the phase boundary for the transition from an orthorhombic to a tetragonal structure. The color gradient represents the smooth evolution of the parameters across the boundary. The red dashed vertical line represents a transition from multiple CDW \textbf{\textit{q}}-vectors to one \textbf{\textit{q}}-vector region.}
\end{figure}

\section{Results and Discussion}

\subsection{Evolution of the crystal structure and the charge density wave}

Powder x-ray diffraction (XRD) spectra along with LeBail fitting for the crushed single crystals with different compositions are shown in \textbf{Figure S1}a. Within the experimental resolution, no secondary phases are detected. \textbf{Figure 1} illustrates the extracted lattice parameters as functions of Sb content at the square-net. We could not prepare any CeSb$_{x}$Te$_{2-x-\delta}$ crystal with $x>$0.91. Therefore, the parameters for CeSbTe are obtained from an earlier report \cite{Schoop3}. Several other parameters from Ref. \cite{Schoop3} are also used throughout this work in order to put the results in context. CeSbTe crystallizes in the ZrSiS-type tetragonal structure (spacegroup $P4/nmm$). Our data for CeSb$_{0.91}$Te$_{0.91}$ are similar to those for ideal stoichiometric CeSbTe \cite{Schoop3}. With decrease in the Sb content, the distorted crystal structure is better described using the orthorhombic spacegroup $Pmmn$. This also explains the orthorhombic structure observed by Wang $et$ $al$. \cite{Wang} and Lv $et$ $al$. \cite{Lv2} for ``CeSbTe'' crystals, which had a significant vacancy at the Sb-site. From our analysis, we identified a boundary around $x$=0.70 for the tetragonal to orthorhombic phase transition. We note that the orthorhombic structure has also been reported in few other members of the $Ln$SbTe-family \cite{DiMasi, Singha2, Lei2}. The lattice parameters undergo a smooth transition across this phase boundary. Two local maxima at $x$=0.91 and 0.34 are observed for both lattice constants $c$ and $b$, respectively. While the peak in $c$ might indicate a deviation from ideal stoichiometry for CeSbTe crystals in Ref. \cite{Schoop3}, the maximum in $b$ represents a phase transition from multiple to one CDW modulation vectors, as we will show later. Signatures of satellite peaks in the XRD spectra, which indicate the presence of the CDW, can be seen in all orthorhombic samples.\\

\begin{figure*}
\includegraphics[width=0.7\textwidth]{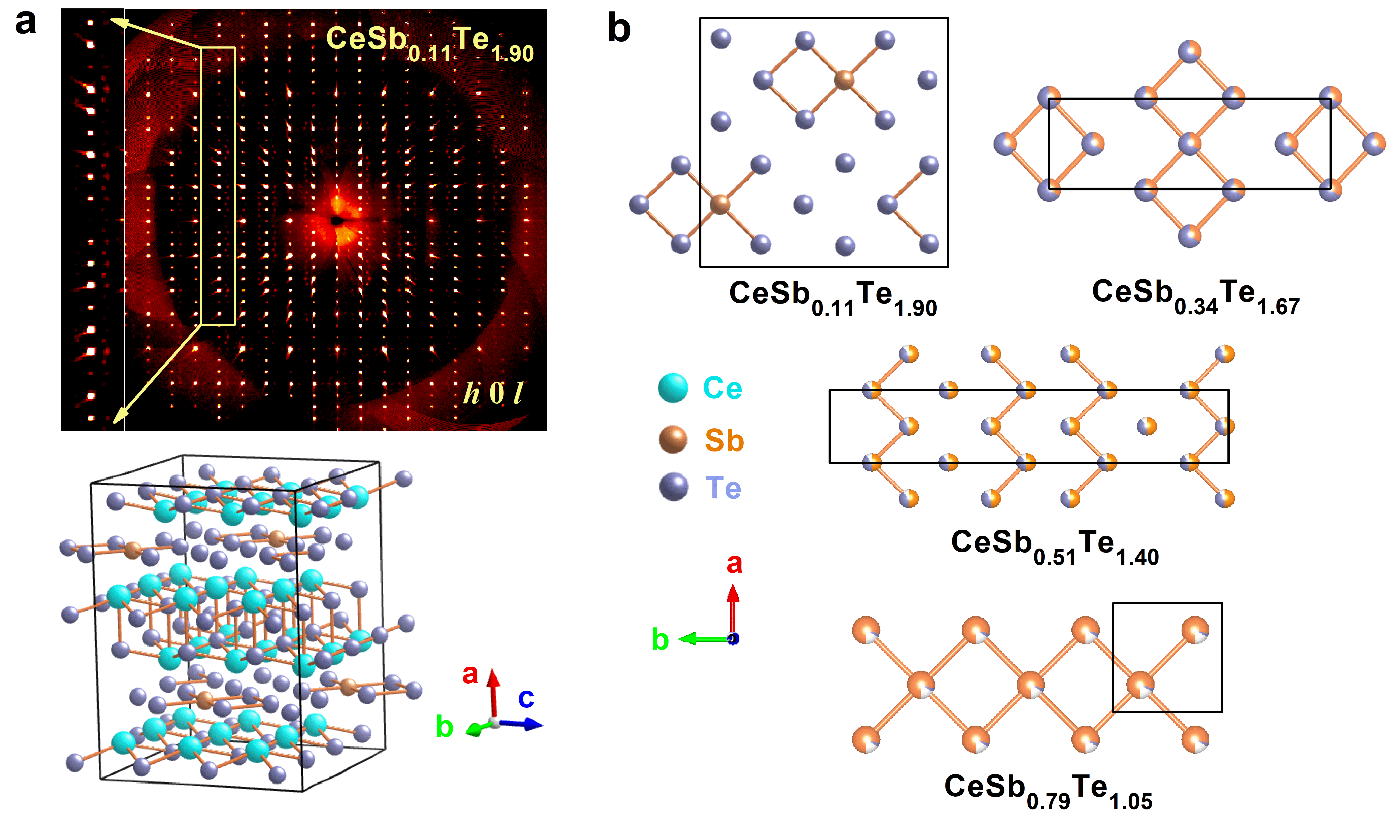}
\caption{Single crystal x-ray diffraction of CeSb$_{x}$Te$_{2-x-\delta}$. a) Precession x-ray diffraction image in the $h$0$l$ plane of a CeSb$_{0.11}$Te$_{1.90}$ single crystal. Enlarged region shows the superlattice reflections generated by three charge density wave (CDW) modulation vectors. The modulated crystal structure is also shown. b) Distorted square-net  motif for compounds with different electron fillings.}
\end{figure*}

The structural evolution of the CDW in CeSb$_{x}$Te$_{2-x-\delta}$ is investigated by single crystal XRD. The presence of a CDW is confirmed by the weak intensity satellite peaks as observed in precession images (\textbf{Figure 2}a and S1b). The solved crystal structures for all the compositions are found to be either commensurately modulated or nearly commensurately modulated (Figure 2a and S1b). In the intermediate Sb-composition range, a single modulation wavevector \textit{\textbf{q}} is observed in the $b$-direction (the longer axis of the $ab$-plane) of the parent structure and it corresponds to a three-fold expansion in CeSb$_{0.34}$Te$_{1.67}$ and a five-fold expansion in the CeSb$_{0.51}$Te$_{1.40}$ unit cell. More complex behavior emerges in the low Sb-composition range. CeSb$_{0.11}$Te$_{1.90}$ exhibits three different \textit{\textbf{q}}-vectors (one within the $ab$-plane, \textit{\textbf{$q_{1}$}}=1/3\textit{\textbf{b}}$^{\ast}$; and two having component along the $c$-axis, \textit{\textbf{$q_{2}$}}=1/3\textit{\textbf{a}}$^{\ast}$+1/3\textit{\textbf{b}}$^{\ast}$+1/2\textit{\textbf{c}}$^{\ast}$ and \textit{\textbf{$q_{3}$}}=1/3\textit{\textbf{a}}$^{\ast}$+1/3\textit{\textbf{b}}$^{\ast}$-1/2\textit{\textbf{c}}$^{\ast}$, where \textit{\textbf{a}}$^{\ast}$, \textit{\textbf{b}}$^{\ast}$, and \textit{\textbf{c}}$^{\ast}$ are the reciprocal lattice vectors) corresponding to multiple distinct CDW orderings, which result in an overall 3$\times$3$\times$2 expansion of the parent cell. This continuously evolving CDW leads to significant distortions of the Sb/Te square-net, yielding a diversity of bonding motifs at the hypervalent square-net (Figure 2b) as Sb-composition corresponds to further a localization of bonding in the structure. Considering a maximum bond distance of 3.1 {\AA} in the undistorted square-net, the Sb/Te square-net distorts into patterns containing zig-zag chains and isolated atoms in CeSb$_{0.51}$Te$_{1.40}$, fused 4-member rings in CeSb$_{0.34}$Te$_{1.67}$, and fish-like functionalized 4-member rings with isolated atoms in CeSb$_{0.11}$Te$_{1.90}$. As evident from Figure S1b, no CDW ordering is observed in tetragonal CeSb$_{0.79}$Te$_{1.05}$. The results of the structural solution for all compositions are summarized in \textbf{Table S1}.\\

\begin{figure}
\includegraphics[width=0.45\textwidth]{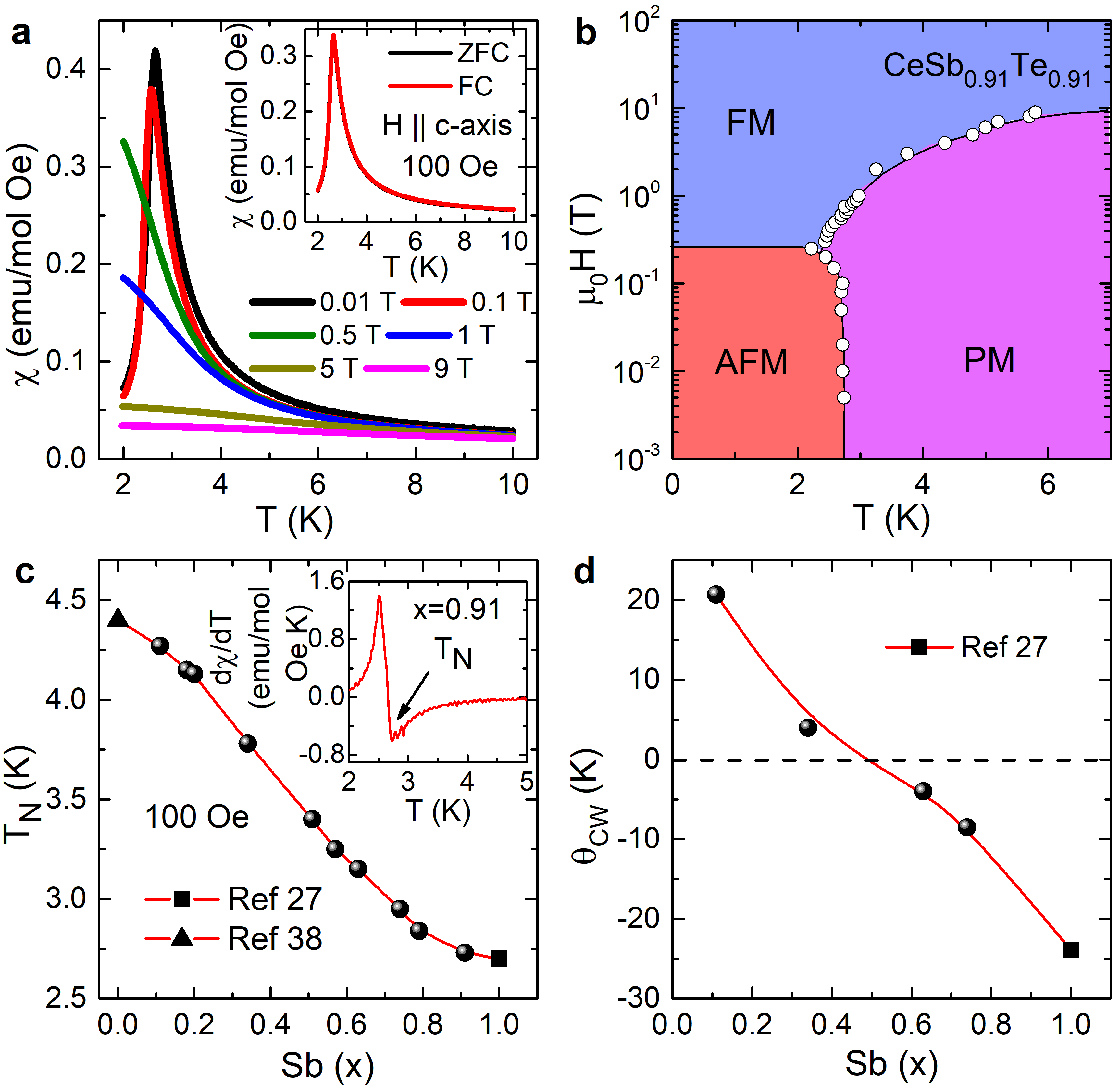}
\caption{Magnetization measurements of CeSb$_{x}$Te$_{2-x-\delta}$ single crystals. a) Temperature dependent susceptibility ($\chi$) for CeSb$_{0.91}$Te$_{0.91}$ at different magnetic fields applied along the crystallographic $c$-axis. Inset shows the zero-field-cooled (ZFC) and field-cooled (FC) magnetization curves at 100 Oe. b) Phase diagram of CeSb$_{0.91}$Te$_{0.91}$, constructed from the magnetization measurements. c) Evolution of the antiferromagnetic N\'eel temperature ($T_{N}$) with Sb-content. Inset shows a typical plot of the first order derivative of $\chi$, which is used to extract $T_{N}$. d) Doping dependence of the Curie-Weiss temperature.}
\end{figure}

\subsection{Magnetic phase diagrams}

In \textbf{Figure 3}a, we have plotted the temperature dependence of the magnetic susceptibility ($\chi$) of tetragonal (unmodulated) CeSb$_{0.91}$Te$_{0.91}$ for different magnetic field strengths, applied along the crystallographic $c$-axis. In general the results are very similar to those observed in CeSbTe \cite{Schoop3}, showing that within the tetragonal region, the magnetic properties do not change drastically with doping. Just like CeSbTe, \newline
CeSb$_{0.91}$Te$_{0.91}$ is antiferromagnetic at low applied fields with N\'eel temperature ($T_{N}$) $\sim$2.73 K. From the powder neutron diffraction measurements, the exact magnetic structure of CeSbTe was found to be type-A AFM, where moments of Ce$^{3+}$ ions are aligned parallel to each other within a layer along the $ab$-plane and these layers are stacked along the $c$-axis with AFM coupling between two consecutive layers \cite{Schoop3}. With an applied magnetic field along the $c$-axis, the moments of all the Ce-layers start to align parallel to the field, resulting in susceptibility curves similar to a ferromagnetic (completely spin-polarized) state. From the transition temperature and field, we have constructed a phase diagram for CeSb$_{0.91}$Te$_{0.91}$ in Figure 3b, showing three distinct magnetization regions - AFM, completely spin polarized [ferromagnetic (FM)], and paramagnetic (PM). In CeSbTe, a group theory analysis revealed that it is possible to realize different types of topological band structures including Dirac (both type I and II) and Weyl nodes as well as higher order band crossings (threefold and eightfold degenerate), depending on the nature and orientation of the magnetic ordering \cite{Schoop3}. Therefore, by controlling the temperature and magnetic field, it is possible to tune the electronic band structure of this material. Here, we are introducing the electron filling at the square-net as an additional tuning parameter. Figure 3c illustrates the evolution of $T_{N}$ with Sb-content. The inset shows the first order derivative of the $\chi$($T$) curve, used for extracting $T_{N}$, for a representative crystal. We observe that $T_{N}$ increases significantly with decreasing Sb-composition, confirming that the AFM state becomes more stable. For $x$=0.11, $T_{N}$ is found to be 4.27 K, which also agrees well with the transition temperature (4.4 K) of CeTe$_{2}$ \cite{Jung}, the other terminal compound. To extract the Curie-Weiss temperature ($\theta_{CW}$), we fit the inverse susceptibility $\chi^{-1}$ ($T$) curve in the PM region using Curie-Weiss law [$\chi$=$\frac{N_{A}\mu_{eff}^{2}}{3k_{B}(T-\theta_{CW})}$] for some representative compositions (\textbf{Figure S2}). While $\theta_{CW}$ is negative for higher Sb-compositions as expected for AFM ordering, it changes sign below $x\sim$0.50 indicating some FM component [Figure 3d]. This might not be unexpected as we note that the magnetization in CeTe$_{2}$ has been reported to be quite complex. There are conflicting reports of a ferrimagnetic ground state with positive $\theta_{CW}$ \cite{Park, Stowe} as well as a long range AFM ground state coexisting with a short range FM ordering \cite{Jung}. From Curie-Weiss fitting for all the compositions, the effective moment ($\mu_{eff}$) is calculated to be close to the moment (2.54 $\mu_{B}$) of the free Ce$^{3+}$ ion. Nevertheless, some deviation from the theoretical value is observed, which is possibly due to the strong crystal electric field effect \cite{Tsuchida}.\\

\begin{figure*}
\includegraphics[width=0.8\textwidth]{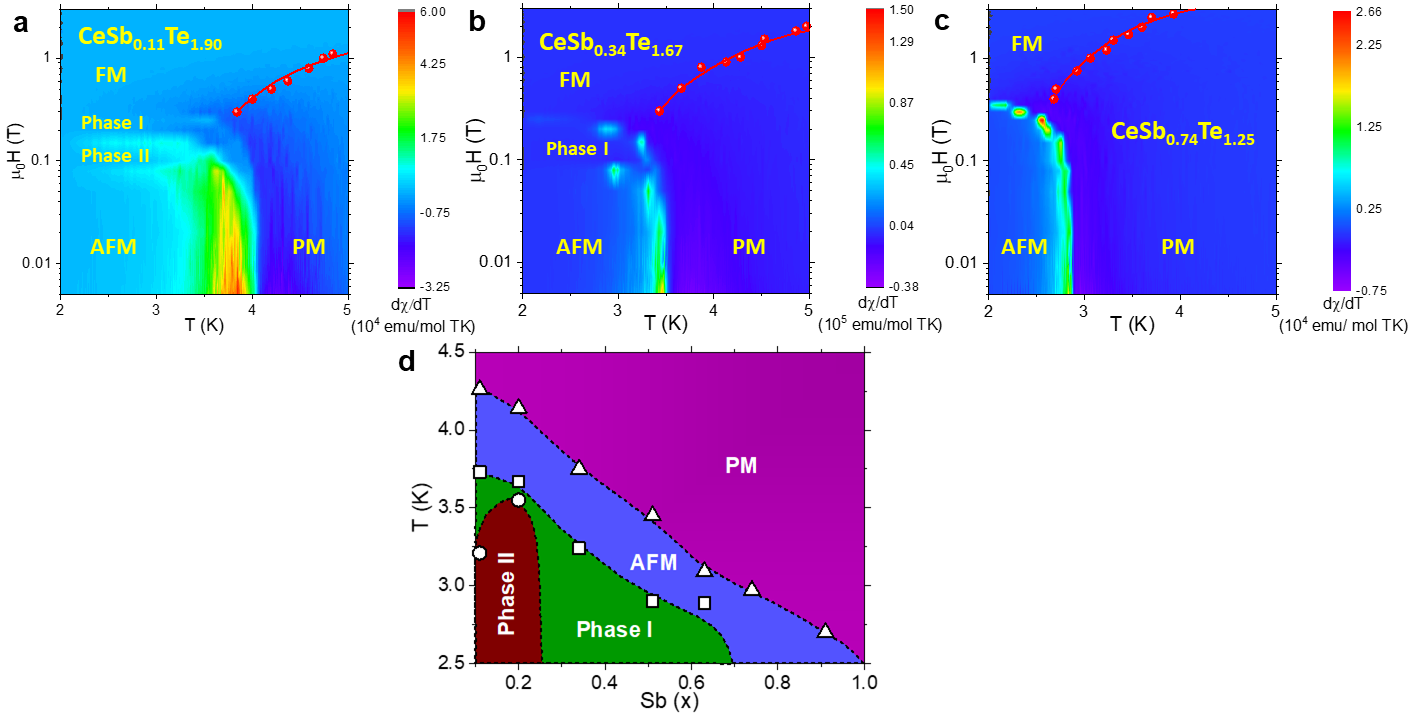}
\caption{Magnetic phase diagrams of CeSb$_{x}$Te$_{2-x-\delta}$. a)-c) Magnetic phase diagram for three different Sb-content samples, showing three distinct magnetization regimes for CeSb$_{x}$Te$_{2-x-\delta}$. The discrete points and red curve, obtained from the magnetization data, represent the boundary between FM and PM states. d) Doping dependent phase diagram.}
\end{figure*}

To get the complete picture, we have constructed magnetic phase diagrams from the first order derivative of $\chi$ ($T$) curves with the field parallel to the $c$-axis for the entire electron filling range (\textbf{Figure 4}a-c and \textbf{S3}a). From the results, it is clear that there are three distinct regimes, represented by three compositions in Figure 4a-c. For $x$=0.11 (orthorhombic with three \textbf{\textit{q}}-vectors), in addition to the AFM and FM states that are similar to the ones in tetragonal CeSbTe, we observe two new phases, namely `Phase I' and `Phase II'. Among these, Phase II is a field-induced state, whereas the boundary for Phase I extends towards lower fields and becomes indistinguishable from the AFM phase boundary. To identify the true magnetic ground state, we have performed a zero-field heat capacity ($C_{P}$) measurement on a CeSb$_{0.11}$Te$_{1.90}$ crystal (\textbf{Figure S4}a). From the enlarged view of the low-temperature region in Figure S4b, we can clearly see an additional peak in $C_{P}$ adjacent to the AFM transition, confirming that Phase I coexists even at zero-field. Upon application of a magnetic field, this second peak becomes more prominent as also evident from the emergence of an explicit boundary with field in the magnetic phase diagram. Both of these peaks in $C_{P}$ are suppressed completely at 0.5 T, indicating a fully spin-polarized state. With increasing Sb-content at $x$=0.34 (orthorhombic with three-fold modulation along the $b$-axis), Phase II disappears, whereas Phase I still remains. For tetragonal CeSb$_{0.74}$Te$_{1.25}$, however, the phase diagram becomes much simpler and closely related to CeSbTe \cite{Schoop3}. In Figure 4d, we have plotted the doping dependent phase diagram. From this figure, we can correlate the boundaries for both new magnetic phases with the structural transitions. Phase II only appears in presence of multiple CDW orderings, whereas Phase I only exists in orthorhombic structures. The previously published magnetization data for CeTe$_{2}$ suggests that Phase I probably corresponds to short range FM ordering \cite{Jung}. On the other hand, Phase II represents a more complex magnetic state as we discuss below.\\

Contrary to the complex magnetic structure for field along the $c$-axis, the phase diagrams for all the compositions turn out to be quite simple, when the magnetic field is applied along the $ab$-plane (Figure S3b). In this configuration, only three states have been observed (AFM, FM, and PM), similar to the $c$-axis phase diagram of CeSbTe \cite{Schoop3}.\\

\textbf{Figure 5}a shows the magnetization ($M$) as a function of magnetic field ($H$ $\parallel$ $c$-axis) for tetragonal \newline
CeSb$_{0.91}$Te$_{0.91}$ at different temperatures both below and above $T_{N}$. To compare, the $M$($H$) curve at 2 K for H $\perp$ $c$-axis is also plotted in the same graph. The material reaches the FM state at lower field, when H $\parallel$ $c$-axis, confirming that it is the easy axis of magnetization. Similar to in CeSbTe, we resolve the `spin-flip' (H $\parallel$ $c$-axis) and `spin-flop' (H $\perp$ $c$-axis) transitions [Figure 5b: arrows illustrate the spin arrangements between two consecutive Ce-layers], as well as strong magnetocrystalline anisotropy.\\

\begin{figure}
\includegraphics[width=0.5\textwidth]{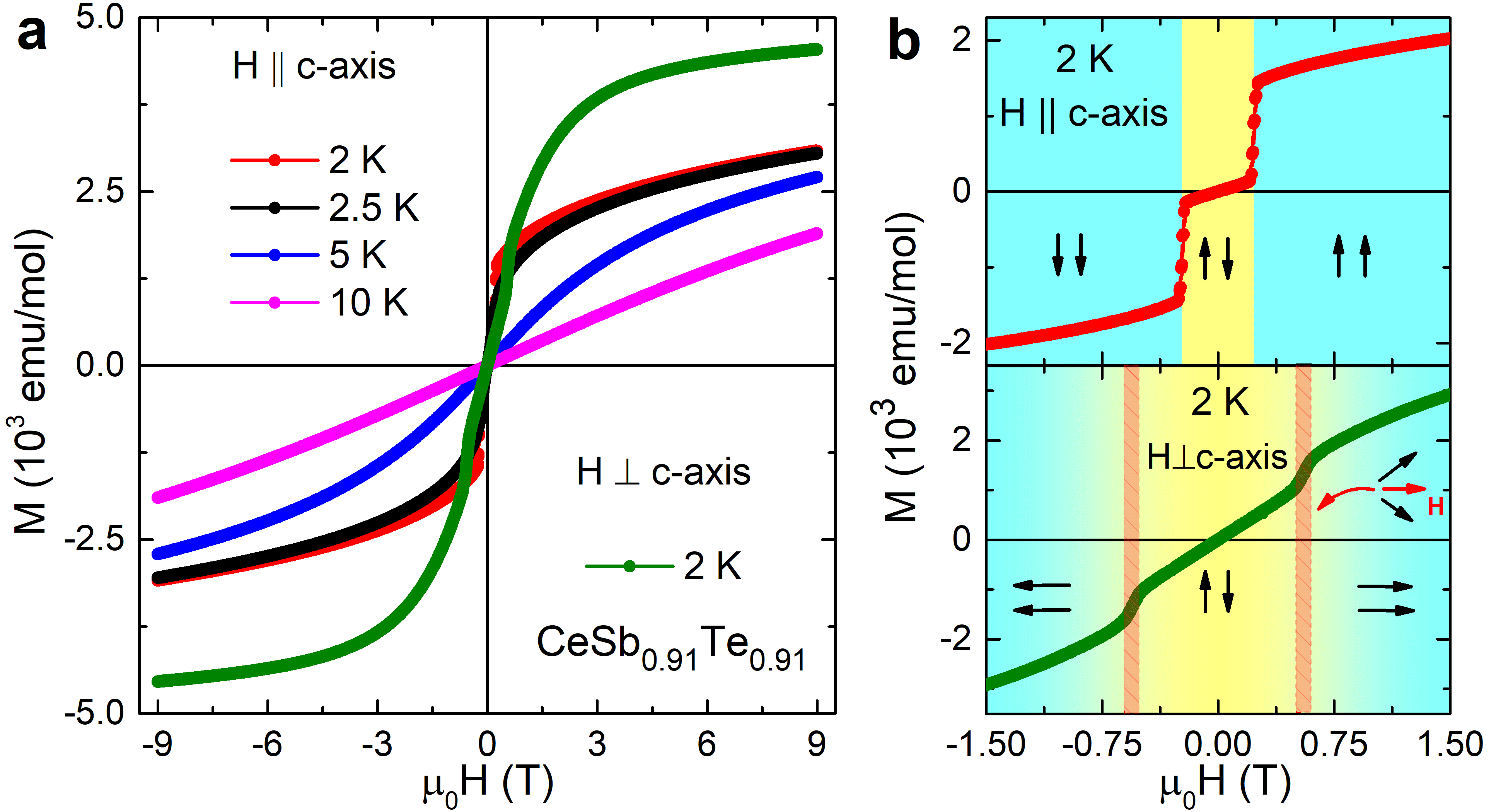}
\caption{Field dependence of magnetization for CeSb$_{0.91}$Te$_{0.91}$. a) Magnetization ($M$) at different temperatures as a function of field ($H$), applied along the $c$-axis. For comparison, the $M$($H$) curve at 2 K for H $\perp$ $c$-axis is also plotted. b) The low-field region of the $M$($H$) curves, showing `spin-flip' (upper panel) and `spin-flop' (lower panel) transitions for two different applied field directions. The arrows illustrate the spin arrangement for two consecutive Ce-layers.}
\end{figure}

In \textbf{Figure S5}a, we have compared the $M$($H$) curves of all Sb-compositions for two measurement configurations. It is evident that the magnetocrystalline anisotropy changes systematically with the electron filling, whereas the direction of spin-flip and spin-flop transitions remain unaltered, indicating that the spins are still aligned along the $c$-axis. For $x$=0.11, the nature of anisotropy is completely opposite to that for $x$=0.91. This suggests that the electron filling at the Sb square-net position continuously tunes the AFM exchange interaction between two consecutive Ce-layers (nearest neighbors). This is indeed expected, as the square-net motif in CeSb$_{x}$Te$_{2-x-\delta}$ provides the conduction electrons, required for Ruderman-Kittel-Kasuya-Yosida (RKKY) interaction between localized Ce$^{3+}$ moments at two different layers. We find that the anisotropy reverses at a composition close to $x$=0.50, where the Curie-Weiss temperature also changes its sign indicating predominantly the AFM ground state for $x\geq$0.50. It is possible that below this critical Sb-content, with higher electron filling, the FM interaction between two next nearest neighbor layers starts to dominate and contributes to a positive Curie-Weiss temperature.\\

\begin{figure*}
\includegraphics[width=0.65\textwidth]{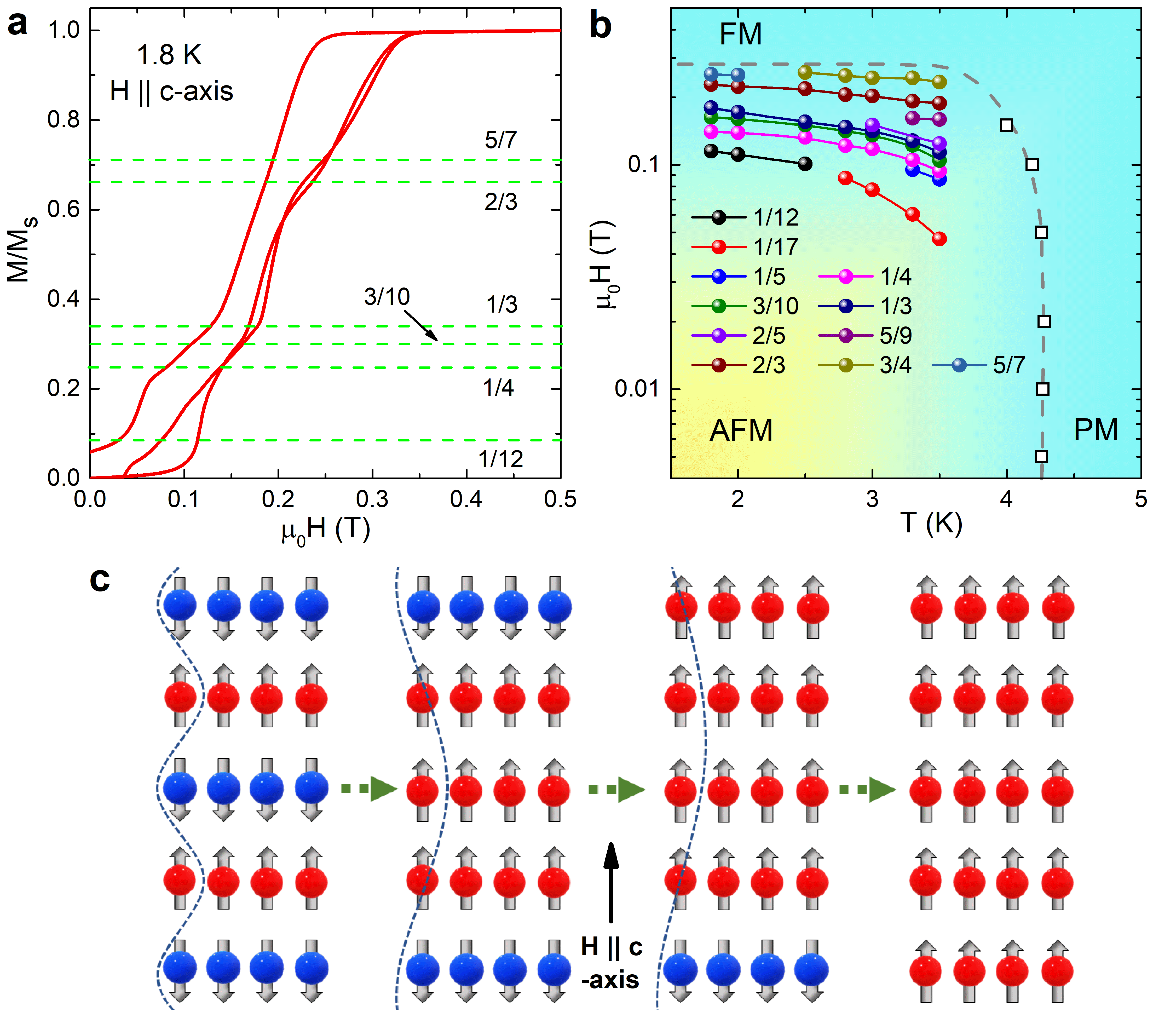}
\caption{Devil's staircase in magnetization of CeSb$_{0.11}$Te$_{1.90}$. a) Normalized magnetization as a function of field, revealing fractionally quantized plateaus. b) Magnetic phase diagram for plateaus corresponding to different quantum numbers. c) Schematic showing the magnetic field-tunable spin wave, responsible for devil's staircase structure.}
\end{figure*}

\subsection{Fractionally quantized magnetization plateaus}

An even more surprising magnetic structure appears at high electron filling. Figure S5b shows the low-field region (-0.5 T$\leq H \leq$0.5 T) of the $M$($H$) curves for samples with $x\leq$0.51, when H $\parallel$ $c$-axis. For $x$=0.11 and 0.20 (multiple \textbf{\textit{q}}-vector region), we observe cascades of metamagnetic transitions, represented by a series of plateaus and steps in magnetization, leading up to the spin-flip transition. In addition, there is prominent hysteresis between the increasing and decreasing-field measurements, confirming the presence of a FM component. The metamagnetic transitions significantly weaken and then completely disappear at $x\geq$0.34. In \textbf{Figure 6}a, we have plotted the $M$($H$) curve at 1.8 K for CeSb$_{0.11}$Te$_{1.90}$, normalized by the saturation magnetization ($M_{S}$). Remarkably, the plateaus are found to be locked to rational fractions identical to the quantized Hall resistivity in the fractional quantum Hall effect (FQHE). Quantized magnetization plateaus have been previously observed in low-dimensional magnets \cite{Shiramura, Kageyama} and magnetically frustrated systems \cite{Ono, Nikuni}, forming a structure called ``Devil's staircase'' as each step consists of infinite number of steps under magnification, and so forth \cite{Bak}. We note that such complex magnetic structure has also been reported in CeSbSe with a tetragonal structure \cite{Chen2}. Analogous to the FQHE, a devil's staircase originates from an energy gap in the many body excitation spectra in a compound due to the translational symmetry breaking \cite{Oshikawa, Momoi, Oshikawa2, Momoi2}. The plateau appears when a commensurability condition between lattice wave vector and localized excitation is satisfied. For quantum spin systems, this commensurability condition is found to be $n$($S$-$m$)=integer, where $n$, $S$, and $m$ are number of spins in the unit cell, the magnitude of spin, and magnetization per site, respectively \cite{Oshikawa}. In principle, this staircase structure can be observed for different physical properties in presence of two coupled waves \cite{Bak}. A continuous variation in frequency (wavelength) of one wave then drives the frequency of other wave such that they go through regimes of phase-locked (plateau) and non-phase-locked (steps) states.\\

By repeating the magnetization measurements for CeSb$_{0.11}$Te$_{1.90}$ at several temperatures (\textbf{Figure S6}), we have tracked the evolution of the plateaus corresponding to different fractions. Using these results, in Figure 6b, a phase diagram is constructed. We conclude that this staircase structure is a magnetic field induced state and only appears within the AFM phase. Interestingly, it is also sensitive to the crystallographic directions, as no metamagnetic transition is observed, when the field is applied perpendicular to the $c$-axis (\textbf{Figure S7}). Moreover, the doping range ($x$=0.11 and 0.20), where this complex state is observed also suggests that it corresponds to the Phase II in the magnetic phase diagram. We note that only within this region, there are multiple CDW modulation wave-vectors, two with components along the $c$-axis and the other one residing within the $ab$-plane. Below this electron filling, the CDW wave-vectors along the $c$-axis disappear. All these features help us to construct a microscopic picture of the origin of the devil's staircase structure in CeSb$_{x}$Te$_{2-x-\delta}$ as illustrated in Figure 6c.\\

The AFM ordered ($\uparrow\downarrow\uparrow\downarrow$) Ce$^{3+}$-layers in CeSb$_{x}$Te$_{2-x-\delta}$ form a spin wave along the $c$-axis. With application of a magnetic field perpendicular to the layers, the down spins try to flip in order to align with the field. However, under a low enough field strength, this transition does not occur at all down-spin layers simultaneously. Instead, different layers undergo spin-flip transitions one after another, thus, effectively creating a spin-wave with continuously field-tunable wavelengths. As the magnetic exchange interaction between these layers is mediated by conduction electrons through the RKKY interaction, the spin-wave is already coupled to any modulation in the electron density along the $c$-axis. Moreover, the $c$-axis component of the two CDW wave-vectors is commensurate with the lattice, hence also with the spin-wave modulation-vector. So, by controlling the magnetic field, these coupled waves can be driven to a series of sequential phase-locked and non-phase-locked states. Once all the spins are aligned along the magnetic field (fully spin-polarized state), there is no longer a spin-wave structure and hence, no more steps are observed. On the other hand, for $x>$0.2, the absence of a \textbf{\textit{q}}-vector along the $c$-axis causes the steps to disappear. We note that in cubic CeSb, a continuous modulation of the spin-wave with both temperature and field, and its commensurability with lattice wave-vector, have been confirmed to be the origin of the devil's staircase in $M$($T$, $H$) \cite{Mignod, Boehm}. In particular, the ``1/3 plateau'' is observed when the layers are arranged in a $\uparrow\uparrow\downarrow$ configuration. Interestingly, two extremely weakly interacting spin-excitations with one of them having analogous properties to astrophysical ``dark matter'', have recently been observed within the ``1/3 plateau'' of CeSb \cite{LaBarre}, showing the possibility of exploring new quantum states in the fractional magnetization plateaus. In the case of CeSb$_{x}$Te$_{2-x-\delta}$, we observe another signature of the spin-wave in the magnetic entropy ($S_{m}$), calculated from the heat capacity data (Figure S4c). For $x$=0.51, $S_{m}$ reaches a saturation value $Rln$2, expected for localized Ce$^{3+}$ moments, whereas it is smaller for $x$=0.11, indicating a gapped spin excitation spectrum \cite{Wiebe, Ramirez}.\\

\begin{figure*}
\includegraphics[width=0.7\textwidth]{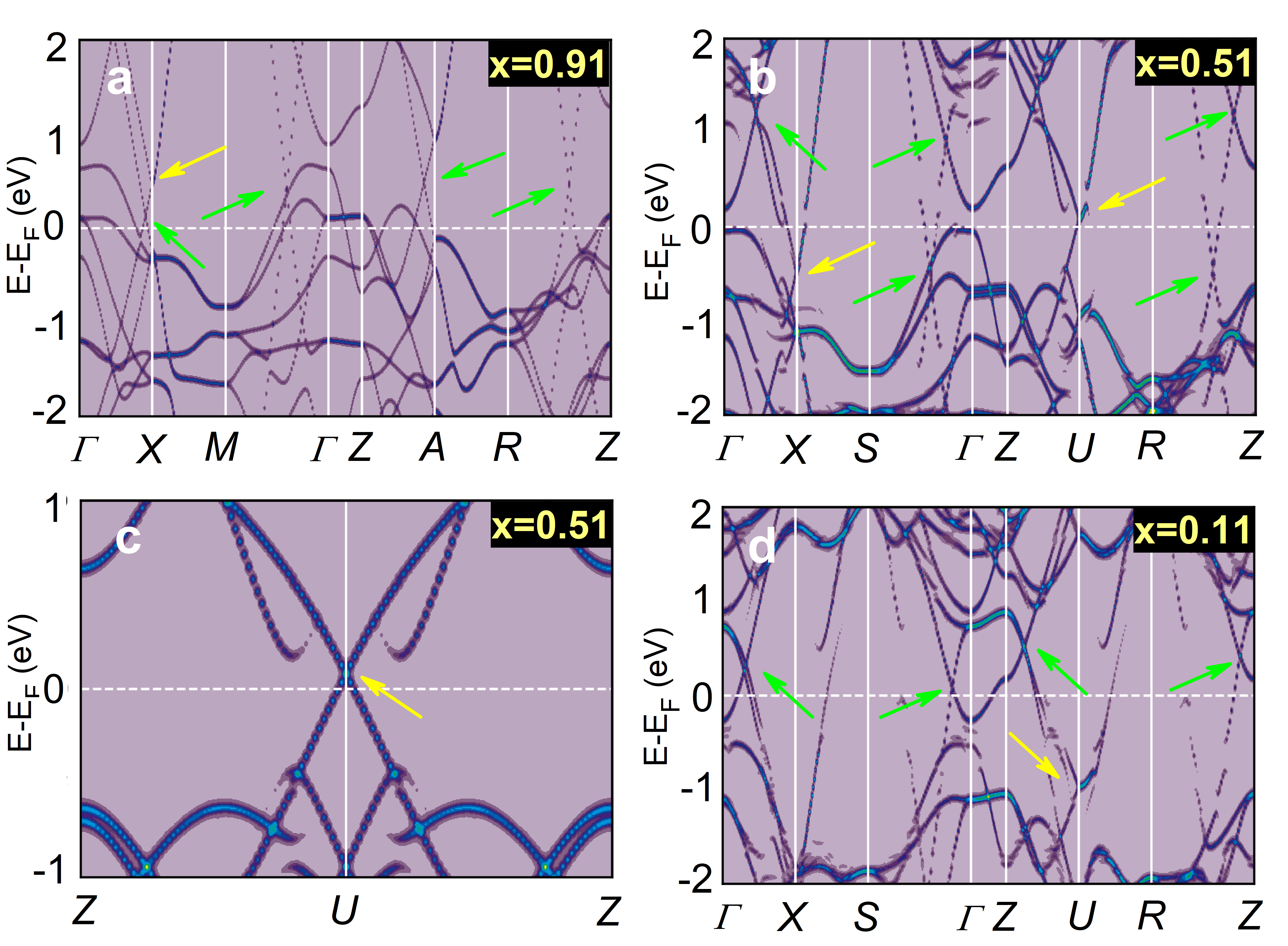}
\caption{Electronic band structure of CeSb$_{x}$Te$_{2-x-\delta}$. Results of the band structure calculations for a) tetragonal CeSb$_{0.91}$Te$_{0.91}$ and b) orthorhombic CeSb$_{0.51}$Te$_{1.40}$ with one \textbf{\textit{q}}-vector. c) Ideal non-symmorphic Dirac cone in CeSb$_{0.51}$Te$_{1.40}$ at the Fermi energy. d) Electronic band structure of orthorhombic CeSb$_{0.11}$Te$_{1.90}$ with multiple \textbf{\textit{q}}-vectors. The yellow and green arrows show the positions of the non-symmorphic Dirac node and nodal-line crossings, respectively.}
\end{figure*}

\subsection{Topological properties of the electronic band structure}

Next, we investigate the topological nature of CeSb$_{x}$Te$_{2-x-\delta}$. As mentioned above, CeSbTe has been shown to be of interest for several distinct topological semimetal phases, which can be modulated by an applied magnetic field. In fact, non-trivial topological states are either predicted or confirmed in different members of the $Ln$SbTe-family \cite{Singha2, Hosen, Pandey, Yue}. Recently, the influence of electron count on the topological band structure has been studied in GdSb$_{x}$Te$_{2-x-\delta}$, where it was shown that an idealized nonsymmorphic Dirac semimetal can be achieved in GdSb$_{0.46}$Te$_{1.48}$ \cite{Lei}. So, it is of interest to also study the band structures of CeSb$_{x}$Te$_{2-x-\delta}$ with respect on their topological nature. We have performed first-principle calculations for three different compositions, representing different regimes of the structural phase diagram (orthorhombic structure with multiple \textbf{\textit{q}}-vectors, one \textbf{\textit{q}}-vector, and tetragonal structure). In the tetragonal state (CeSb$_{0.91}$Te$_{0.91}$), the band structure (\textbf{Figure 7}a) unsurprisingly closely resembles the one of CeSbTe \cite{Schoop3}, with only a slight difference in electron filling. For the single \textbf{\textit{q}}-vector range, the band structure of CeSb$_{0.51}$Te$_{1.40}$ is shown in Figure 7b, which features a five-fold supercell along the $b$-axis. Orthorhombic CeSb$_{0.51}$Te$_{1.40}$ is isostructural to GdSb$_{0.46}$Te$_{1.48}$ \cite{Lei} and analogous to this compound, it also hosts a close to ideal nonsymmorphic Dirac semimetal band structure. Nonsymmorphic symmetry protected gapless Dirac nodes reside near the Fermi energy ($E_{F}$) at high symmetry points $X$ and $U$ (yellow arrows), whereas almost all other band crossings gapped out by the CDW and do not contribute to the low-energy transport response (Figure 7c). While it seems that some bands still cross $E_{F}$ (as CDW band gap is smaller than in its Gd-counterpart), there are small gaps in their spectra exactly at the Fermi level. This is interesting for two reasons: (i) it suggests that the mechanism reported in Ref. \cite{Lei} can be universally applied to \textit{Ln}Sb$_{x}$Te$_{2-x-\delta}$ phases that are in the single \textbf{\textit{q}}-vector region with \textbf{\textit{q}}$\approx$0.20\textit{\textbf{b}}$^{\ast}$. (ii) it provides a second idealized nonsymmorphic Dirac semimetal, which features completely different magnetic properties. Most importantly, in CeSb$_{x}$Te$_{2-x-\delta}$ the spins are aligned along the $c$-axis, while GdSb$_{x}$Te$_{2-x-\delta}$ they are aligned in plane (perpendicular to the $c$-axis) \cite{Lei3}. This provides an additional degree of tunability for magnetically induced new topological phases and opens the door to investigate further \textit{Ln}Sb$_{x}$Te$_{2-x-\delta}$ phases that should also show an idealized Dirac semimetal band structure.

Finally, in Figure 7d, we show the band structure of CeSb$_{0.11}$Te$_{1.90}$, which has multiple \textbf{\textit{q}}-vectors. In this compound, due to the infinite number of magnetic phases that arise from the devil's staircase, the tunability of topological phases becomes extremely rich. As can be seen in Figure 7d, the band structure of CeSb$_{0.11}$Te$_{1.90}$ features both nonsymorphically enforced Dirac crossing (yellow arrow) as well as nodal-line crossings (green arrows). Especially, a nodal-line crossing resides exactly at $E_{F}$ along $S$-$\Gamma$. In addition, there are multiple nodal-line crossings just above $E_{F}$, which should be easily accessible by fine tuning of the electron filling at the Te-site. Future studies could further investigate the interplay of the topological band structure and devil's staircase magnetism in the multiple \textbf{\textit{q}}-vector region of \textit{Ln}Sb$_{x}$Te$_{2-x-\delta}$.\\

\section{Conclusion}

To conclude, we have probed the detailed structural and magnetic properties of  topological semimetals CeSb$_{x}$Te$_{2-x-\delta}$ as a function of electron filling at the Sb-square-net site. We found that the crystal structure increasingly transforms from a tetragonal structure at ideal stoichiometry to an orthorhombic one with electron filling. This distorted orthorhombic structure enables formation of CDWs with continuously tunable modulation wave-vectors. More interestingly, two distinct regimes have been observed, hosting either one or multiple CDWs along different crystallographic axes. Complete solutions of the modulated structures have been obtained, which also reveal the unique distortions of the square-net motif at different electron filling regions. Contrary to the simple AFM ordering in CeSbTe, the magnetic phase diagram of CeSb$_{x}$Te$_{2-x-\delta}$ emerges to be quite rich and evolves continuously with chemical substitution. Specifically, at higher electron filling, we show that a complex interplay of CDW and a collective spin-excitation leads to a series of fractionally quantized magnetization plateaus, which offers the potential of realizing new quantum phases in these compounds. The results of our electronic band structure calculations confirm that CeSb$_{0.51}$Te$_{1.40}$ is an ideal magnetic Dirac semimetal with a non-symmorphic Dirac node at the Fermi energy, while almost all other bands are gapped out. Thus we present a unique template material, where the topological band structure can be controlled by electron filling (tuning chemical potential), CDWs (gapping out non-essential band crossings), or magnetic field (changing the magnetic ordering). This also encourages future studies to design new topological states in CeSb$_{x}$Te$_{2-x-\delta}$ and other members of the $Ln$SbTe-family.\\

\section{Experimental Section}

\subsection{Single crystal growth and determination of stoichiometry}
Single crystals of CeSb$_{x}$Te$_{2-x-\delta}$ were grown by chemical vapor transport. Stoichiometric amounts of high purity Ce (Sigma Aldrich 99.9\%), Sb (Alfa Aesar 99.999\%), and Te (Alfa Aesar 99.9999\%) along with a few milligrams of iodine (Sigma Aldrich 99.999\%) were placed into a quartz tube. The tube was then evacuated, sealed, and put into a gradient furnace. The furnace was heated such that the hot end of the quartz tube, containing the materials, remained at 950$^{\circ}$ C and the other end at 850$^{\circ}$ C for 7 days. After cooling, the crystals were mechanically extracted from the cold end of the tube. The elemental composition of the crystals was determined by energy dispersive x-ray spectroscopy (EDX) in a Verios 460 scanning electron microscope, operating at 15 keV and equipped with an Oxford EDX detector.\\

\subsection{X-ray diffraction and magnetic measurements}
The powder XRD measurements were performed on a STOE STADI P diffractometer operating in transmission geometry using Mo-$K_{\alpha}$ ($\lambda$=0.71073 {\AA}) source. The single crystals were crushed into powder in an argon-filled glovebox and sealed in glass capillary to use for XRD experiments. The XRD spectra was analyzed by LeBail fitting using FULLPROF software package.

The single crystal XRD data were collected at 250(1) K with either a Bruker Kappa Apex2 CCD diffractometer or a Bruker D8 VENTURE equipped with a PHOTON CMOS detector, using \newline
graphite-monochromatized Mo-$K_{\alpha}$ radiation. Raw data were corrected for background, polarization, and Lorentz factors as well as a multiscan absorption correction was applied. Structure solution was carried out via either intrinsic phasing as implemented in the ShelXT program or via charge-flipping as implemented in SUPERFLIP. Commensurately modulated phases for CeSb$_{0.11}$Te$_{1.90}$ and CeSb$_{0.34}$Te$_{1.67}$; and tetragonal structure of CeSb$_{0.79}$Te$_{1.05}$ were able to be indexed fully and could be refined in conventional three-dimensional space groups, using the ShelXL least-squares refinement package in the Olex2 program. In the modulated phase CeSb$_{0.51}$Te$_{1.40}$, the satellite peaks refined to a slightly incommensurate modulation vector \textit{\textbf{q}}=0.201\textit{\textbf{b}}$^{\ast}$. Refinement was thus carried out using the superspace approach, where the displacive distortion of atomic positions is expressed by a periodic modulation function, yielding a 3+1 dimensional space group \cite{Smaalen}. Refinements with the superspace approach were carried out in JANA2006. For the refinement, the modulation vector was rounded to a commensurate \textit{\textbf{q}}=1/5\textit{\textbf{b}}$^{\ast}$ without any significant changes to the fitting compared to a fully incommensurate treatment. Determination of Sb/Te occupancy and ordering within the square-net is limited by near-identical scattering power due to their closeness in atomic number ($Z$=51 and 52, respectively) \cite{Hammond}. At laboratory-accessible x-ray wavelengths, the two species are indistinguishable if present on the same crystallographic site. Thus, for all refinements, atomic occupancies within the Sb/Te square-net were constrained to the stoichiometry derived from EDX. Assignment of ordering in CeSb$_{0.11}$Te$_{1.90}$ was based on the chemical bonding intuitions preferring multiple bonds to Sb.

The magnetic measurements were performed using the vibrating sample magnetometer (VSM) option of a physical property measurement system (Quantum Design).\\

\subsection{Band structure calculations}
Density functional theory (DFT) calculations were performed in VASP \textit{v}5.4.4 \cite{kresse1994abinitio,kresse1996efficient,kresse1996efficiency} using the PBE functional \cite{perdew1996generalized}. Localization of the Ce $f$-orbitals was corrected by applying a Hubbard potential of $U$=6eV \cite{dudarev1998electron}, as in previous work on tetragonal CeSbTe \cite{Schoop3}. PAW potentials \cite{blochl1994projector,kresse1999from} were chosen based on the \textit{v}5.2 recommendations. Simulations approximating CeSb$_{0.91}$Te$_{0.91}$, CeSb$_{0.51}$Te$_{1.40}$, and CeSb$_{0.11}$Te$_{1.90}$ were performed on the tetragonal CeSb$_{0.91}$Te$_{0.91}$ unit cell with full occupancy and supercells of 1$\times$5$\times$1 Ce$_{10}$Sb$_{10}$Te$_{10}$ and 3$\times$3$\times$2 Ce$_{36}$Te$_{68}$Sb$_4$ with Fermi levels adjusted based on the electron counts of the true experimental cells. Calculations employed a plane wave energy cutoff of 400\,eV and a $k$-mesh density, $\ell=30$ (corresponding to 7$\times$7$\times$3 and 7$\times$1$\times$3 and 2$\times$2$\times$2 $\Gamma$-centered $k$-meshes for the tetragonal CeSb$_{0.91}$Te$_{0.91}$ subcell and the CeSb$_{0.51}$Te$_{1.40}$ and CeSb$_{0.11}$Te$_{1.90}$ supercells, respectively). Unfolded spectral functions for the supercells in the subcell BZ were calculated using the method of Popescu and Zunger \cite{popescu2012extracting} in \textsc{VaspBandUnfolding}. The $\Gamma$-$X$ high symmetry line in CeSb$_{0.51}$Te$_{1.40}$ (the only elongated cell for which this is ambiguous) was chosen to lie along the distortion axis. Crystal structures were visualized with \textsc{VESTA} \cite{momma2011vesta}. Spin-orbit coupling effects were not incorporated. See additional computational details in the supporting material.\\

\medskip
\textbf{Acknowledgements} \par
This research was supported by the Princeton Center for Complex Materials, a National Science Foundation (NSF)-MRSEC program (DMR-2011750). Additional support for property characterization was provided by the Air Force Office of Scientific Research (AFOSR), grant number FA9550-20-1-0282. The authors acknowledge the use of Princeton’s Imaging and Analysis Center, which is partially supported by the Princeton Center for Complex Materials. Work at UC Santa Barbara was supported by the National Science Foundation though the Q-AMASE-i Quantum Foundry, (DMR-1906325). We acknowledge use of the shared computing facilities of the Center for Scientific Computing at UC Santa Barbara, supported by NSF CNS-1725797, and the NSF MRSEC at UC Santa Barbara, NSF DMR-1720256. S.M.L.T. has been supported by the NSF Graduate Research Fellowship Program under Grant no. DGE-1650114.  Any opinions, findings, and conclusions or recommendations expressed in this material are those of the authors and do not necessarily reflect the views of the NSF.

\medskip
\textbf{Conflict of interest} \par
The authors declare no conflict of interest.

\medskip
\textbf{Author contributions} \par
R.S. and L.M.S. initiated the project. R.S. synthesized single crystals and characterized them with input from S.L. R.S. measured and analyzed magnetization data, with input from N.P.O. and L.M.S. T.H.S. performed single crystal x-ray diffraction experiments and solved structures with input from J.F.K. S.M.L.T. performed the band structure calculations. L.M.S. supervised the project. All authors discussed the results and contributed to writing the manuscript.

\newpage

\textbf{Supplementary information: Evolving Devil's staircase magnetization from tunable charge density waves in nonsymmorphic Dirac semimetals}\\

\setcounter{figure}{0}
\textbf{Figure S1}a shows the powder x-ray diffraction spectra of the CeSb$_{x}$Te$_{2-x-\delta}$ crystals. The data are analyzed using LeBail fitting. Within the experimental resolution, no secondary phases are observed.\\

\begin{figure*}
\includegraphics[width=0.8\textwidth]{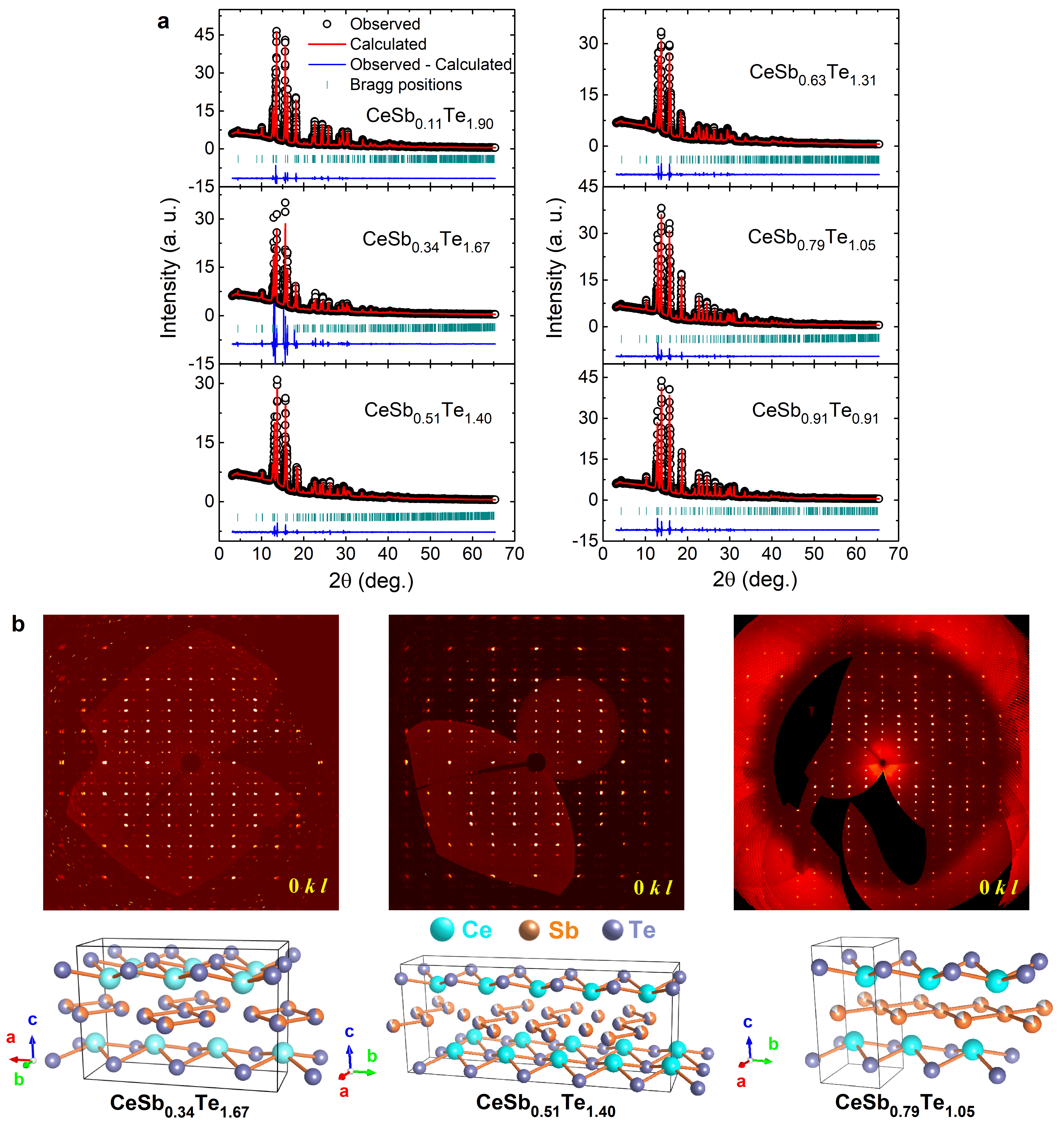}
\renewcommand{\figurename}{Figure S}
\caption{Powder and single crystal x-ray diffraction of CeSb$_{x}$Te$_{2-x-\delta}$. a) Powder x-ray diffraction spectra for different Sb-compositions. The experimental data are analyzed using LeBail fitting. b) Precession diffraction image in the 0$kl$ plane of the single crystals with different electron fillings at the square-net. Superlattice reflections can be clearly seen for $x$=0.34 and 0.51. The corresponding solved crystal structures are also shown.}
\end{figure*}

The single crystal x-ray diffraction results of the CeSb$_{x}$Te$_{2-x-\delta}$ crystals are shown in Figure S1b. The superlattice reflections in precession diffraction images confirm the formation of three-fold and five-fold modulated charge density waves (CDWs) along the $ab$-plane for orthorhombic CeSb$_{0.34}$Te$_{1.67}$ and CeSb$_{0.51}$Te$_{1.40}$, respectively. No signature of a CDW is observed for tetragonal CeSb$_{0.79}$Te$_{1.05}$. The corresponding solved crystal structures are also shown. The results of the structural refinements are summarized in Table S1.\\

\begin{figure*}
\includegraphics[width=0.5\textwidth]{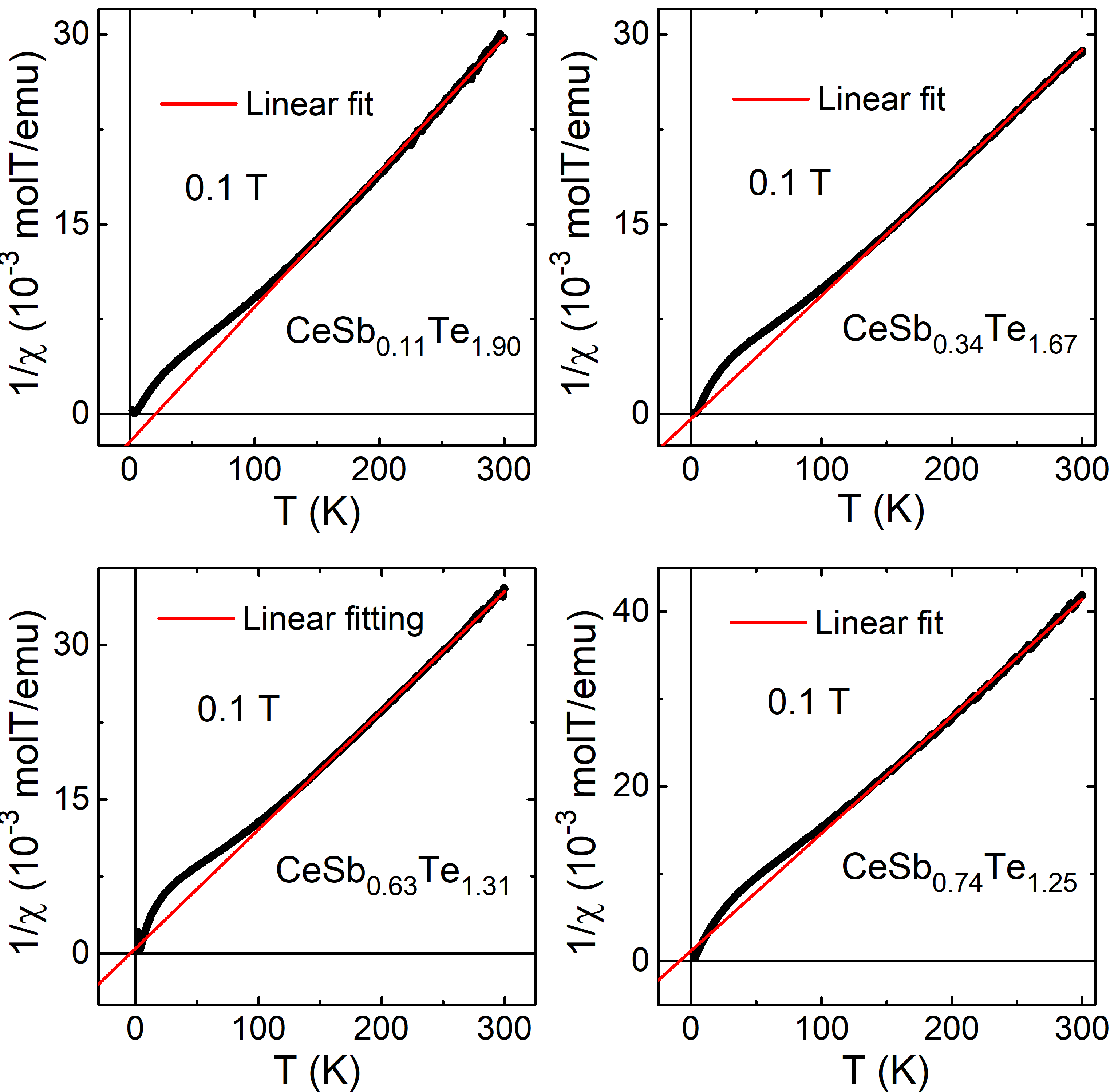}
\renewcommand{\figurename}{Figure S}
\caption{Curie-Weiss fitting. Curie-Weiss fitting of the inverse magnetic susceptibility $\chi^{-1}$($T$) curve in the paramagnetic region for different Sb-compositions.}
\end{figure*}

\textbf{Figure S2} illustrates the Curie-Weiss fits of the inverse magnetic susceptibility $\chi^{-1}$ ($T$) curve in the paramagnetic region for different compositions. The extracted Weiss temperature from the fitting is shown in Figure 3d in main text.\\

\textbf{Figure S3}a shows the magnetic phase diagram of three representative CeSb$_{x}$Te$_{2-x-\delta}$ crystals for magnetic fields applied along the crystallographic $c$-axis. A complex magnetic ground state is observed for higher electron filling. On the other hand, the phase diagram for fields along the $ab$-plane, is found to be quite simple throughout the entire doping range (Figure S3b).\\

\begin{center}
\begin{table*}
\renewcommand{\tablename}{Table S}
\caption{Results of single crystal structural refinement.}
 \begin{tabular}{|c c c c c c|}
 \hline
  & & CeSb$_{0.11}$Te$_{1.90}$ & CeSb$_{0.34}$Te$_{1.67}$ & CeSb$_{0.51}$Te$_{1.40}$ & CeSb$_{0.79}$Te$_{1.05}$\\ [0.5ex]

 \hline\hline
  & Formula weight & 395.95 & 394.61 & 380.86 & 370.29\\
  & Temperature & 250.15 K & 293.0 K & 293 K & 298.15 K\\
  & Wavelength & 0.71073 {\AA} & 0.71073 {\AA} & 0.71073 {\AA} & 0.71073 {\AA}\\
  & Crystal system & Orthorhombic & Orthorhombic & Orthorhombic & Tetragonal\\
  & Space group & $Pnma$ & $Pmmn$ & $Pmmn$(0$\beta$0)00$s$ & $P4/nmm$\\
  & Modulation type & Commensurate & Commensurate & Incommensurate & \\
  & Modulation vectors & \textit{\textbf{$q_{1}$}}=1/3\textit{\textbf{b}}$^{\ast}$ & \textit{\textbf{q}}=1/3\textit{\textbf{b}}$^{\ast}$ & \textit{\textbf{q}}=1/5\textit{\textbf{b}}$^{\ast}$ & \\
  & & \textit{\textbf{$q_{2}$}}=1/3\textit{\textbf{a}}$^{\ast}$+1/3\textit{\textbf{b}}$^{\ast}$+1/2\textit{\textbf{c}}$^{\ast}$ & & & \\
  & & \textit{\textbf{$q_{3}$}}=1/3\textit{\textbf{a}}$^{\ast}$+1/3\textit{\textbf{b}}$^{\ast}$-1/2\textit{\textbf{c}}$^{\ast}$ & & & \\
  & Unit cell dimensions & $a$=13.4641(5) {\AA} & $a$=4.4255(7) {\AA} & $a$=4.3925(1) {\AA} & $a$=4.3853(2) {\AA}\\
  & & $b$=13.4214(5) {\AA} & $b$=13.4684(3) {\AA} & $b$=4.4392(1) {\AA} & $b$=4.3853(2) {\AA}\\
  & & $c$=18.2375(7)  {\AA} & $c$=9.1926(11) {\AA} & $c$=9.3053(3) {\AA} & $c$=9.4000(5) {\AA}\\
  & & $\alpha$=$\beta$=$\gamma$=90$^{\circ}$ & $\alpha$=$\beta$=$\gamma$=90$^{\circ}$ & $\alpha$=$\beta$=$\gamma$=90$^{\circ}$ & $\alpha$=$\beta$=$\gamma$=90$^{\circ}$\\
  & Volume & 3295.6(2) {\AA}$^{3}$ & 547.92(11) {\AA}$^{3}$ & 181.446(8) {\AA}$^{3}$ & 180.770(19) {\AA}$^{3}$\\
  & $Z$ & 36 & 6 & 2 & 2\\
  & Density (calculated) & 7.159 g/cm$^{3}$ & 7.152 g/cm$^{3}$ & 7.1822 g/cm$^{3}$ & 6.803 g/cm$^{3}$\\
  & Absorption coefficient & 27.702 mm$^{-1}$ & 27.635 mm$^{-1}$ & 27.722 mm$^{-1}$ & 26.385 mm$^{-1}$\\
  & F(000) & 5828 & 970 & 314 & 306\\
  & $\theta$ range for data collection & 1.884-29.997$^{\circ}$ & 2.215-44.980$^{\circ}$ & 2.19-72.86$^{\circ}$ & 2.167-36.496$^{\circ}$\\
  & Index ranges & 0$\leq h \leq$25 & -26$\leq h \leq$26 & -11$\leq h \leq$11 & -7$\leq h \leq$7\\
  & & -18$\leq k \leq$18 & -8$\leq k \leq$8 & -12$\leq k \leq$11 & -7$\leq k \leq$7\\
  & & -18$\leq l \leq$18 & -18$\leq l \leq$18 & -23$\leq l \leq$24 & -15$\leq l \leq$15\\
  & & & & -1$\leq m \leq$1 & \\
  & Reflections collected & 18530 & 61201 & 33837 & 8258\\
  & Independent reflections & 4988 & 2524 & 11239 & 309\\
  & Completeness to $\theta$=25.242$^{\circ}$ & 100\% & 99.8\% & 99\% ($\theta$=72.86$^{\circ}$) & 100\%\\
  & $R_{int}$ & 0.0231 & 0.1058 & 0.0534 & 0.0545\\
  & Goodness-of-fit & 1.076 & 1.124 & 2.48 & 1.179\\
  & Final $R$ indices [$I>$2$\sigma$($I$)] & $R_{obs}$=0.0596 & $R_{obs}$=0.0519 & $R_{obs}$=0.0718 & $R_{obs}$=0.0386\\
  & & $wR_{obs}$=0.2715 & $wR_{obs}$=0.1461 & $wR_{obs}$=0.0944 & $wR_{obs}$=0.1038\\
  & $R$ indices [all data] & $R_{all}$=0.0879 & $R_{all}$=0.0622 & $R_{all}$=0.0961 & $R_{all}$=0.0386\\
  & & $wR_{all}$=0.3366 & $wR_{all}$=0.1680 & $wR_{all}$=0.0979 & $wR_{all}$=0.1038\\
  & Extinction coefficient & 0.000026(10) & 0.0012(4) & 4130(150) & 0.021(4)\\
  & Largest diff. peak and hole & 4.793 and -6.749 $e${\AA}$^{-3}$ & 6.926 and -11.907 $e${\AA}$^{-3}$ & 12.64 and -14.55 $e${\AA}$^{-3}$ & 2.577 and -3.190 $e${\AA}$^{-3}$\\
 \hline\hline
  & Refinement method: & Full-matrix & & &\\
  & & least-squares on F$^{2}$. & & &\\
  & $R=\frac{\Sigma||F_{o}|-|F_{C}||}{\Sigma|F_{o}|}$ & & & &\\
  & $wR=[\frac{\Sigma[w(|F_{o}|^{2}-|F_{c}|^{2})^{2}]}{\Sigma[w(|F_{o}|^{4})]}]^{1/2}$ & & & &\\
 \hline
\end{tabular}
\end{table*}
\end{center}

\begin{figure*}
\includegraphics[width=0.9\textwidth]{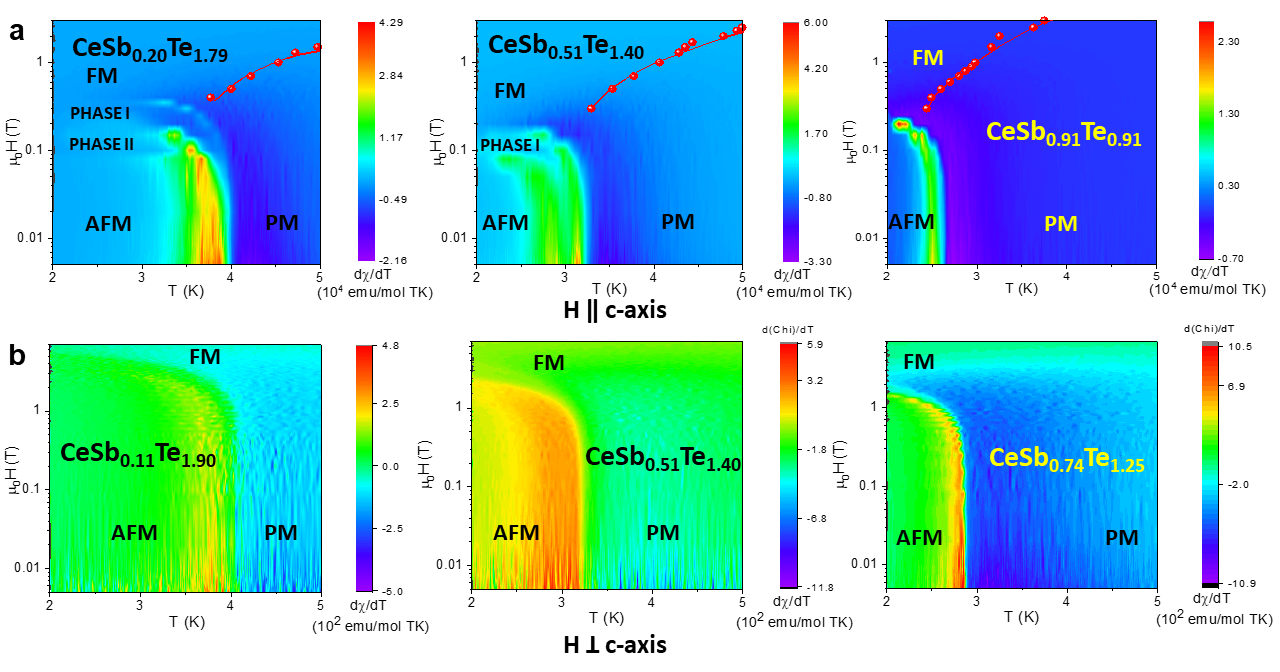}
\renewcommand{\figurename}{Figure S}
\caption{Magnetic phase diagram of CeSb$_{x}$Te$_{2-x-\delta}$. Magnetic phase diagram of different crystals with varying Sb-content for field applied along the crystallographic a) $c$-axis and b) $ab$-plane. The discrete points and red curve, obtained from the magnetization data, represent the boundary between FM and PM states.}
\end{figure*}

\begin{figure*}
\includegraphics[width=0.8\textwidth]{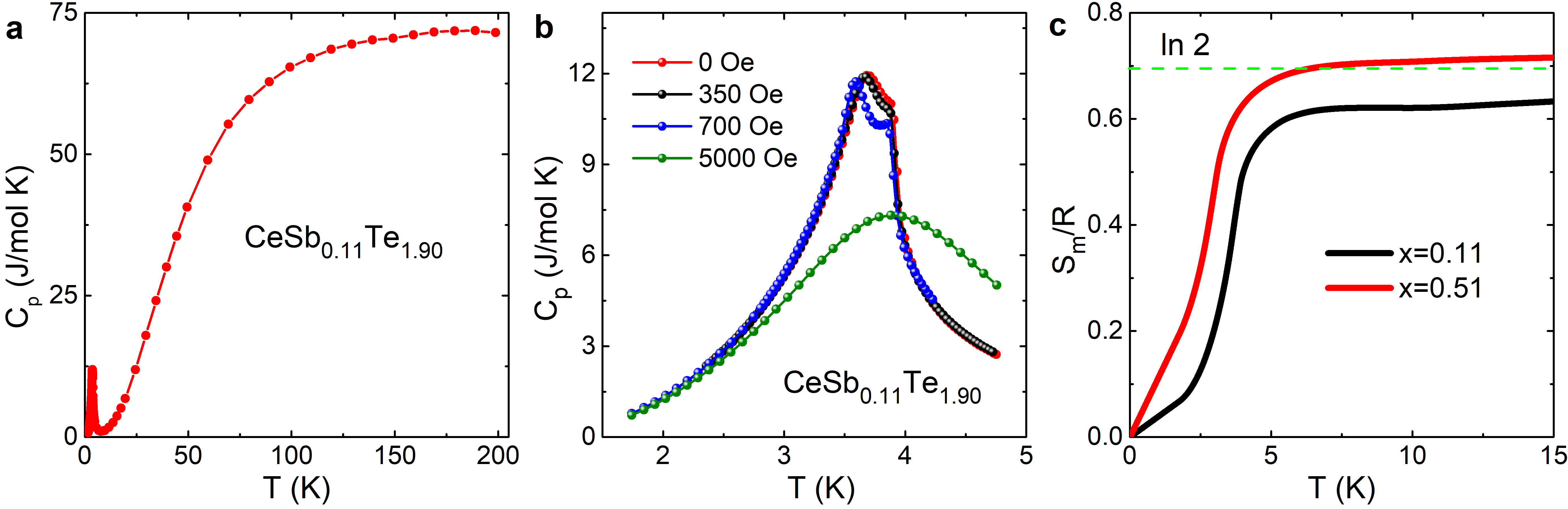}
\renewcommand{\figurename}{Figure S}
\caption{Heat capacity measurements. a) Temperature dependence of the zero-field heat capacity ($C_{P}$) for CeSb$_{0.11}$Te$_{1.90}$. b) Low-temperature region of the heat capacity curve for different magnetic field, applied along the $c$-axis. c) Magnetic entropy ($S_{m}$) calculated from heat capacity data for $x$=0.11 and 0.51.}
\end{figure*}

The zero-field specific heat ($C_{P}$) of CeSb$_{0.11}$Te$_{1.90}$ is plotted in \textbf{Figure S4}a as a function of temperature. The enlarged low-temperature region in Figure S4b clearly shows the presence of two magnetic transitions. With an applied field of 0.5 T, the system reaches a fully spin-polarized state. In order to calculate the magnetic entropy ($S_{m}$), we have fitted the heat capacity data with Debye model, which gives the lattice contribution to $C_{P}$. Subtracting the lattice component from total specific heat and assuming a negligible electronic contribution, magnetic component $C_{m}$ was obtained. The magnetic entropy was calculated using the relation $S_{m}=\int_{0}^{T}\frac{C_{m}}{T}dT$. As the lowest measured temperature was 1.8 K, the specific heat data were interpolated using $C_{P}$=0 at $T$=0 K. This interpolation might introduce a small uncertainty in the calculated value of $S_{m}$. The magnetic entropy for both CeSb$_{0.11}$Te$_{1.90}$ and CeSb$_{0.51}$Te$_{1.40}$ is plotted in Figure S4c.\\

\begin{figure*}
\includegraphics[width=0.9\textwidth]{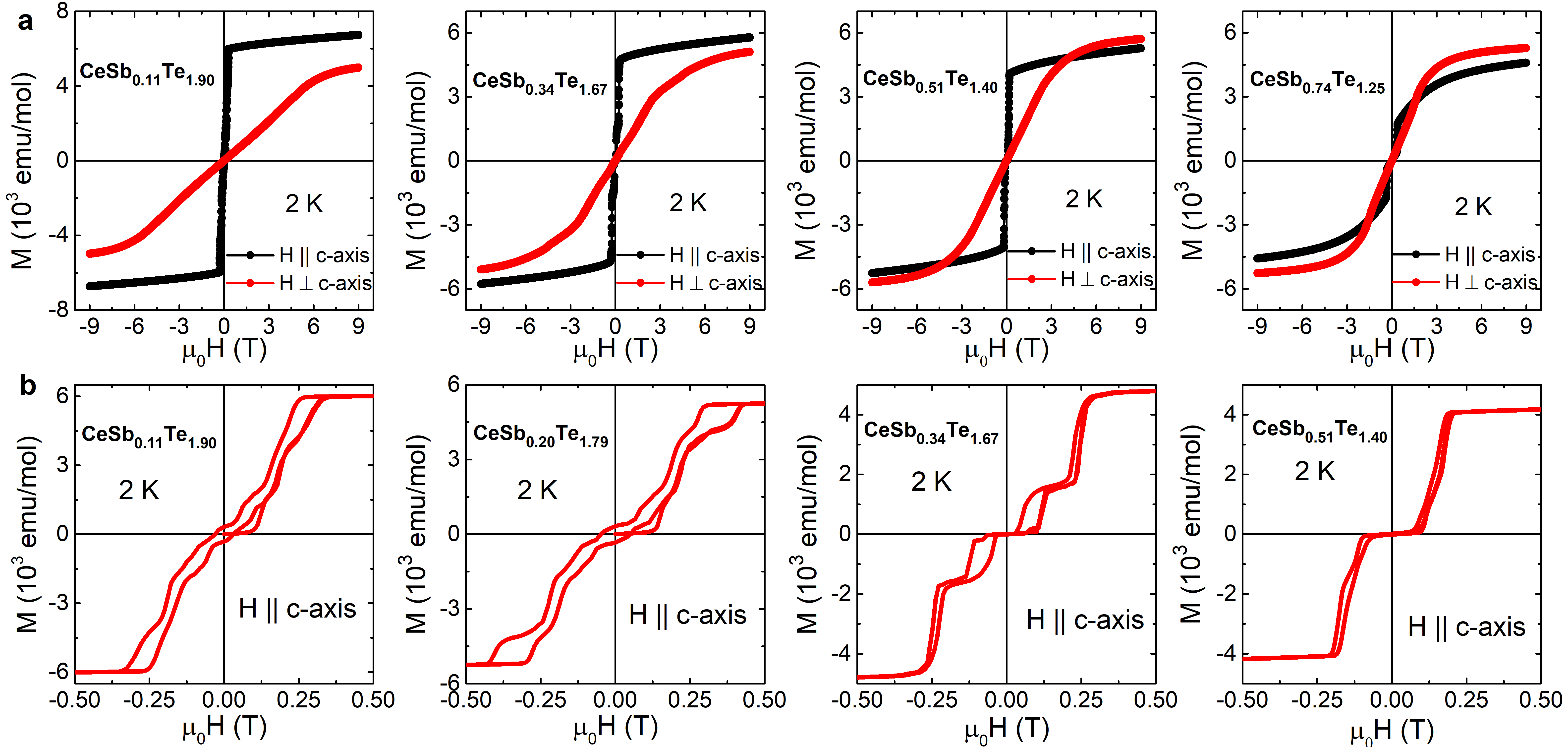}
\renewcommand{\figurename}{Figure S}
\caption{Field dependence of magnetization. a) Magnetization vs. field curves for different CeSb$_{x}$Te$_{2-x-\delta}$ crystals at 2 K with field applied along two mutually perpendicular crystallographic directions, showing reversal of magnetocrystalline anisotropy. b) Low-field region of the magnetization curves for different crystals with field applied along the $c$-axis, showing a series of metamagnetic transitions for higher electron filling.}
\end{figure*}
\textbf{Figure S5}a illustrates the evolution of magnetocrystalline anisotropy with electron filling in CeSb$_{x}$Te$_{2-x-\delta}$. The low-field region of the $M$($H$) curves reveal a series of metamagnetic transitions at higher electron filling (Figure S5b).\\

\begin{figure*}
\includegraphics[width=0.7\textwidth]{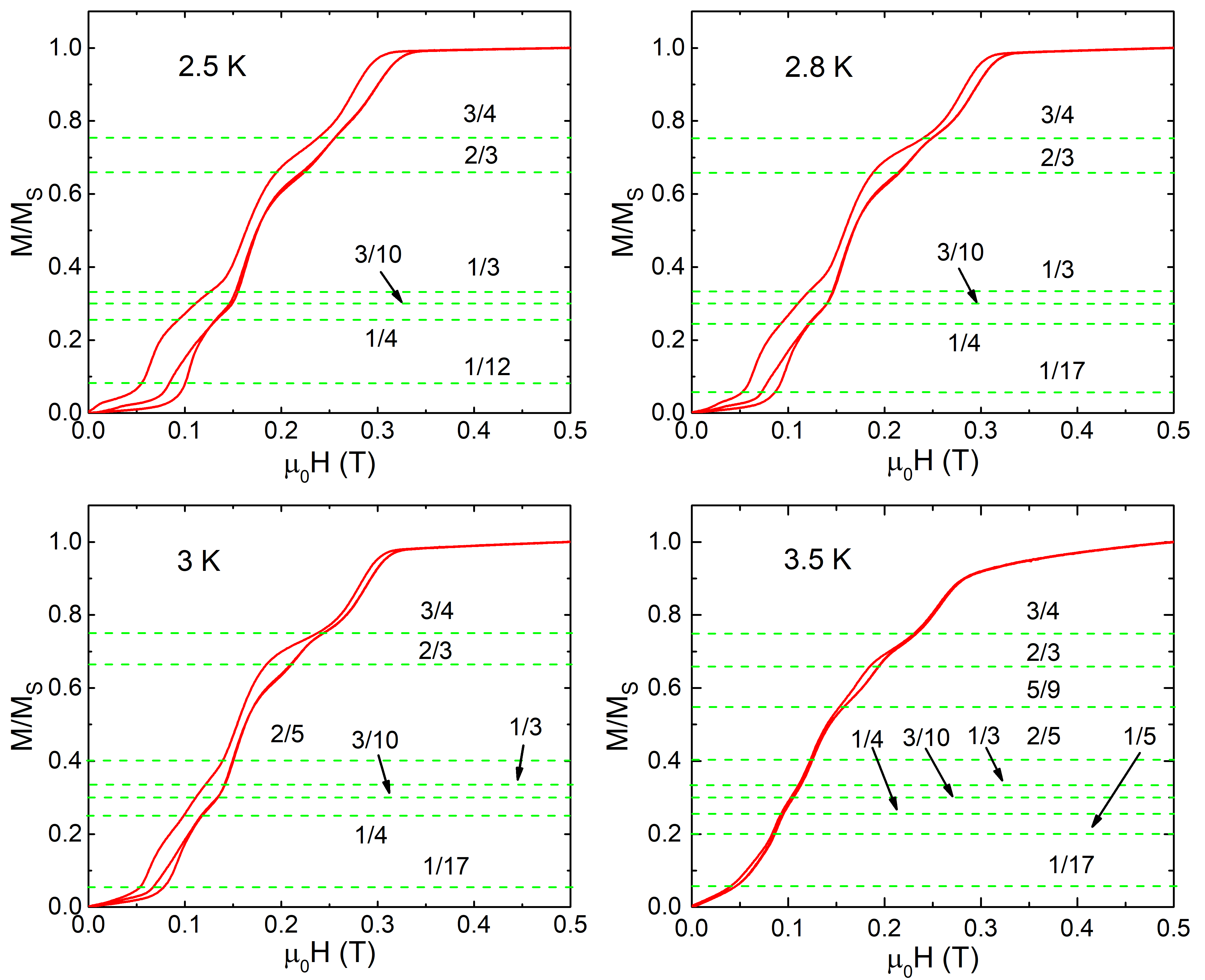}
\renewcommand{\figurename}{Figure S}
\caption{Quantized magnetization plateaus in CeSb$_{0.11}$Te$_{1.90}$. Evolution of magnetization plateaus with temperature for field applied along the $c$-axis.}
\end{figure*}

The quantized magnetization plateaus in CeSb$_{0.11}$Te$_{1.90}$ are shown in \textbf{Figure S6} at different temperatures.

\begin{figure*}
\includegraphics[width=0.45\textwidth]{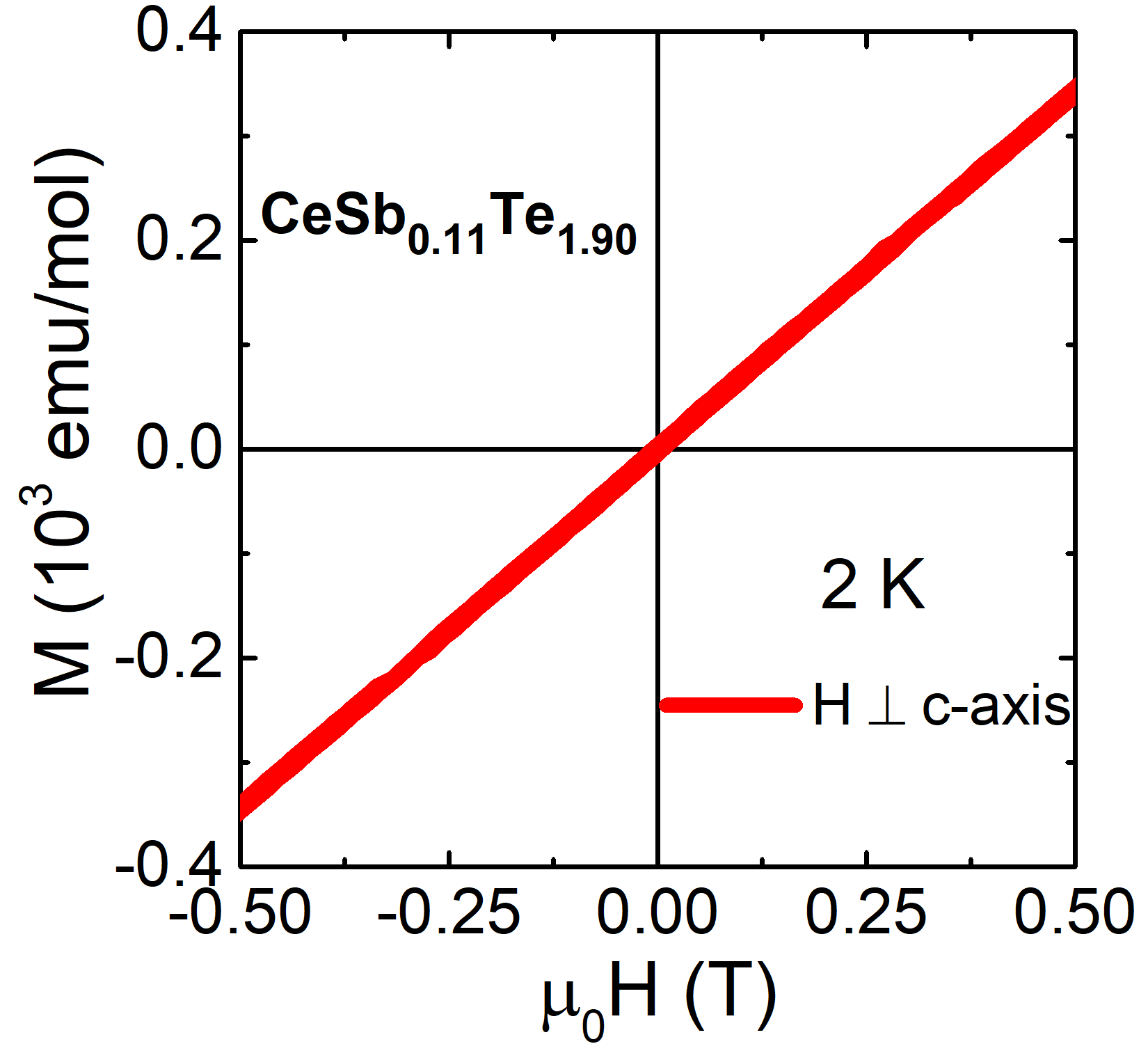}
\renewcommand{\figurename}{Figure S}
\caption{Magnetization curve along the ab-plane for CeSb$_{0.11}$Te$_{1.90}$. Low-field region of the $M$($H$) curve at 2 K for field applied along the $ab$-plane, showing no metamagnetic transition in this measurement configuration.}
\end{figure*}

In \textbf{Figure S7}, the low-field region of the $M$($H$) curve for CeSb$_{0.11}$Te$_{1.90}$ with fields applied along the $ab$-plane, confirms no metamagnetic transition for this measurement configuration.\\

\textbf{Additional computational details:}\\
In \textbf{Figure S8}a, we present the full spin-polarized band structure of CeSb$_{0.91}$Te$_{0.91}$. While compounds in the CeSb$_{x}$Te$_{2-x-\delta}$ family generally have an antiferromagnetic ground state, the spin-splitting near the Fermi level is relatively small, with the result that the band structure is computationally well-approximated by displaying only one spin-channel from the ferromagnetic calculation as we have done in the main body of the text. While simple, this ferromagnetic model has also previously demonstrated good agreement with angle resolved photo emission spectroscopy measurements of the electronic structure of the sister compound GdSb$_{x}$Te$_{2-x-\delta}$ which also undergoes similar Peierls-like distortions \cite{Lei}. In Figure S8b, we present the same band structure with inclusion of spin-orbit coupling effects. Spin-orbit coupling makes only small adjustments to the overall band dispersion. While spin-orbit coupling has not been included in the supercell calculations due to computational expense, it is unlikely to affect our findings about the large Peierls-like band-gapping in these distorted structures.

As a final comment for computationalists interested in these compounds - instead of the true magnetic ground state, metastable magnetic states (with the lowest DFT energy and 0.9-1\,$\mu_{B}$ moments, consistent with experiment) were frequently encountered when studying these systems using DFT. A magnetic energy landscape with many local minima is unsurprising in light of the rich magnetic phase diagram discussed in the main text.

\begin{figure*}
\includegraphics[width=0.45\textwidth]{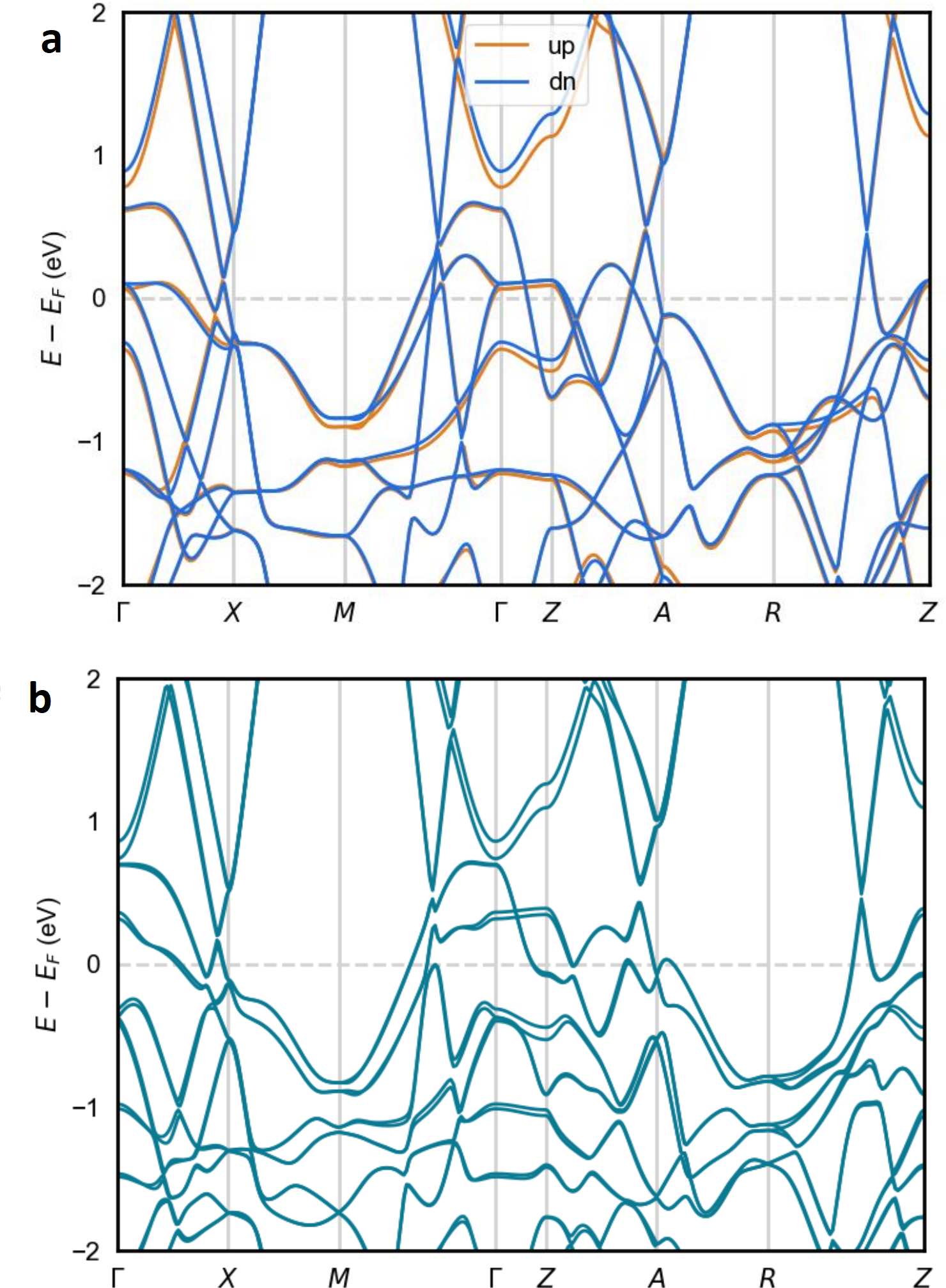}
\renewcommand{\figurename}{Figure S}
\caption{Full band structure of FM CeSb$_{0.91}$Te$_{0.91}$. a) Full spin-polarized calculation with spin-up and spin-down populations in orange and blue, respectively. b) Full band structure after the inclusion of spin-orbit coupling.}
\end{figure*}

\end{document}